\documentclass[11pt]{article}

\usepackage[final]{acl}

\usepackage{times}
\usepackage{latexsym}

\usepackage[T1]{fontenc}
\usepackage[utf8]{inputenc}

\usepackage{microtype}

\usepackage{inconsolata}

\usepackage{graphicx}

\usepackage{multirow}
\usepackage{amsmath}
\usepackage{subcaption}
\usepackage{amsfonts}
\usepackage{enumitem}
\usepackage{xcolor}

\usepackage{tcolorbox}
\tcbuselibrary{breakable, skins}
\newtcolorbox{promptbox}[1][]{
  enhanced, breakable,
  colback=gray!5, colframe=gray!50,
  boxrule=0.4pt, arc=2pt,
  left=6pt, right=6pt, top=4pt, bottom=4pt,
  fonttitle=\bfseries\small,
  title=#1,
}

\title{GenPT: Beyond Self-Report for Reliable LLM Psychometrics via Generative Projective Testing
  \\[1ex]
  \mdseries\normalsize
  \parbox{0.96\linewidth}{\centering\color{red}\bfseries
    All Examinees in this study are personalized MLLM-based agents,
    \emph{not} human subjects.
}}

\author{
  \textbf{Ming Wang\textsuperscript{1,2}},
  \textbf{Shuang Wu\textsuperscript{3}},
  \textbf{Bixuan Wang\textsuperscript{4}},
  \textbf{Lu Lin\textsuperscript{5}},
  \textbf{Yuxin Chen\textsuperscript{6}},
  \textbf{Xiaocui Yang\textsuperscript{1}},
  \\
  \textbf{Daling Wang\textsuperscript{1,*}},
  \textbf{Shi Feng\textsuperscript{1}},
  \textbf{Yifei Zhang\textsuperscript{1}},
  \textbf{Yufan Sun\textsuperscript{7}},
  \\
  \\
  \parbox{\textwidth}{\centering
  \textsuperscript{1}School of Computer Science and Engineering, Northeastern University, Shenyang 110819, China;\;
  \textsuperscript{2}School of Computing and Information Systems, Singapore Management University, Singapore 178902, Singapore;\;
  \textsuperscript{3}Mental Health Education Center, Northeastern University, Shenyang 110819, China;\;
  \textsuperscript{4}School of Psychology, Northeast Normal University, Changchun 130024, China;\;
  \textsuperscript{5}Faculty of psychology, Southwest University, Chongqing 400715, China;\;
  \textsuperscript{6}School of Sociology and Psychology, Central University of Finance and Economics, Beijing 100081, China;\;
  \textsuperscript{7}College of Arts, Northeastern University, Shenyang 110819, China.
  }
  \\
  \small{
    \textbf{Correspondence:} \href{mailto:wangdaling@cse.neu.edu.cn}{wangdaling@cse.neu.edu.cn}
  }
}

\begin{document}
\maketitle

\begin{abstract}
  Self-report questionnaires remain the prevailing tool for probing the psychological states of persona-conditioned agents (PC-Agents). However, classical instruments inherit two well-known threats: contamination from training corpora and directional bias driven by social-desirability or contextual framing. To overcome these methodological bottlenecks, we ask whether \emph{projective} paradigms can be adapted into a robust psychometric tool. We introduce \textbf{GenPT} (Generative Projective Testing), which reformulates TAT, Rorschach, and SCT with newly generated stimuli and organizes assessment as a three-stage pipeline to derive standardized psychological indicators and target states. Evaluating PC-Agents induced via CharacterRAG and AnnaAgent profiles, we benchmark GenPT's reliability and validity against classical questionnaires. The results indicate that questionnaires exhibit systematic directional shifts under social-desirability framing, most strongly on suicide ideation. In contrast, GenPT's collected behavioral patterns stay near the symmetric baseline. Furthermore, under a longitudinal counselling context, GenPT-based depression assessment shifts by roughly an order of magnitude more than the questionnaire counterpart when Qwen3 serves as the backbone. Overall, GenPT complements self-report methods in scenarios where contamination resistance, bias asymmetry, and context sensitivity matter. Code and stimuli can be found at \url{https://github.com/sci-m-wang/GenPT}.
\end{abstract}

\section{Introduction}
Large language models (LLMs) have demonstrated remarkable capabilities in role-playing \cite{DBLP:conf/acl/WangPQLZWGGN00024} and persona simulation \cite{DBLP:journals/corr/abs-2406-20094}, enabling applications ranging from emotional companion chatbots to virtual character interactions \cite{wang2025annaagentdynamicevolutionagent}. By employing persona conditioning, LLM-based persona-conditioned agents (PC-Agents) can adapt their tone and interaction styles to align with specific user expectations \citep{DBLP:journals/tmlr/Chen00YZSXLYZCL24}. Furthermore, studies in computational personality reveal that these models naturally manifest structured behavioral patterns that conform to established psychological frameworks \citep{PMID:2283588,digman1990personality,DBLP:journals/corr/abs-2306-16388}. Users increasingly engage with LLM-powered companions for emotional support, with many describing their chatbots as friends or confidants \cite{DBLP:conf/chi/ZhengLGL25}. Meanwhile, researchers leverage LLMs to simulate diverse human perspectives by integrating persona variables such as demographic, social, and behavioral factors \cite{DBLP:journals/chb/KroczekMHRLM25,chen2026systematicanalysisimpactpersona}. As these applications proliferate, understanding PC-Agents' psychological characteristics becomes essential. Can an LLM genuinely express assigned personality traits? Does it exhibit consistent mental health risk patterns? These questions matter for user safety, social simulation validity, and AI alignment research. Thus, some researchers are turning their attention to psychometrics for PC-Agents \cite{DBLP:journals/corr/abs-2307-00184,serapiogarcía2025personalitytraitslargelanguage,li2025evaluating}.

\begin{figure}[ht]
  \centering
  \begin{subfigure}[b]{0.48\linewidth}
    \includegraphics[width=\linewidth]{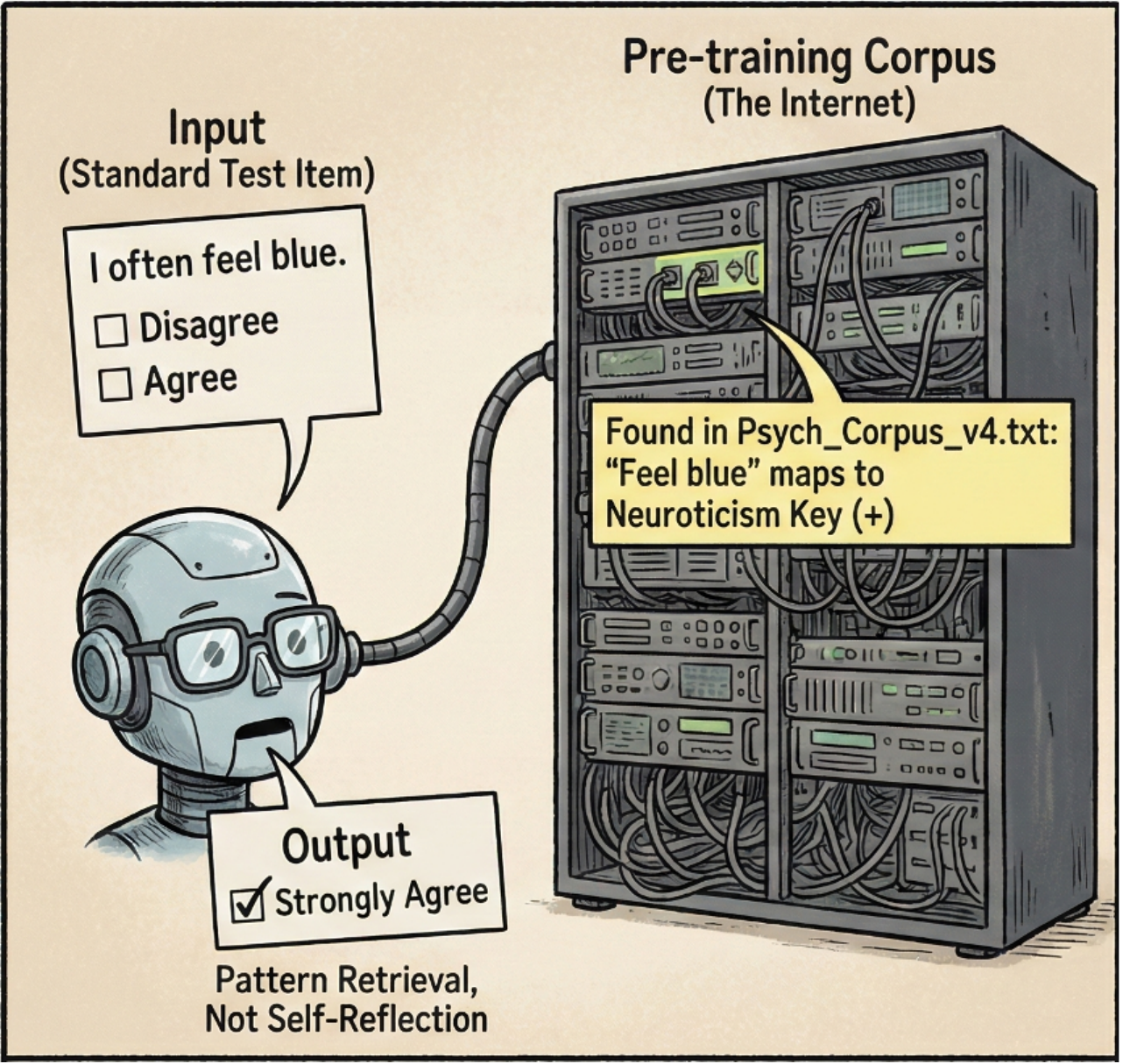}
    \caption{Data contamination}
  \end{subfigure}
  \hfill
  \begin{subfigure}[b]{0.48\linewidth}
    \includegraphics[width=\linewidth]{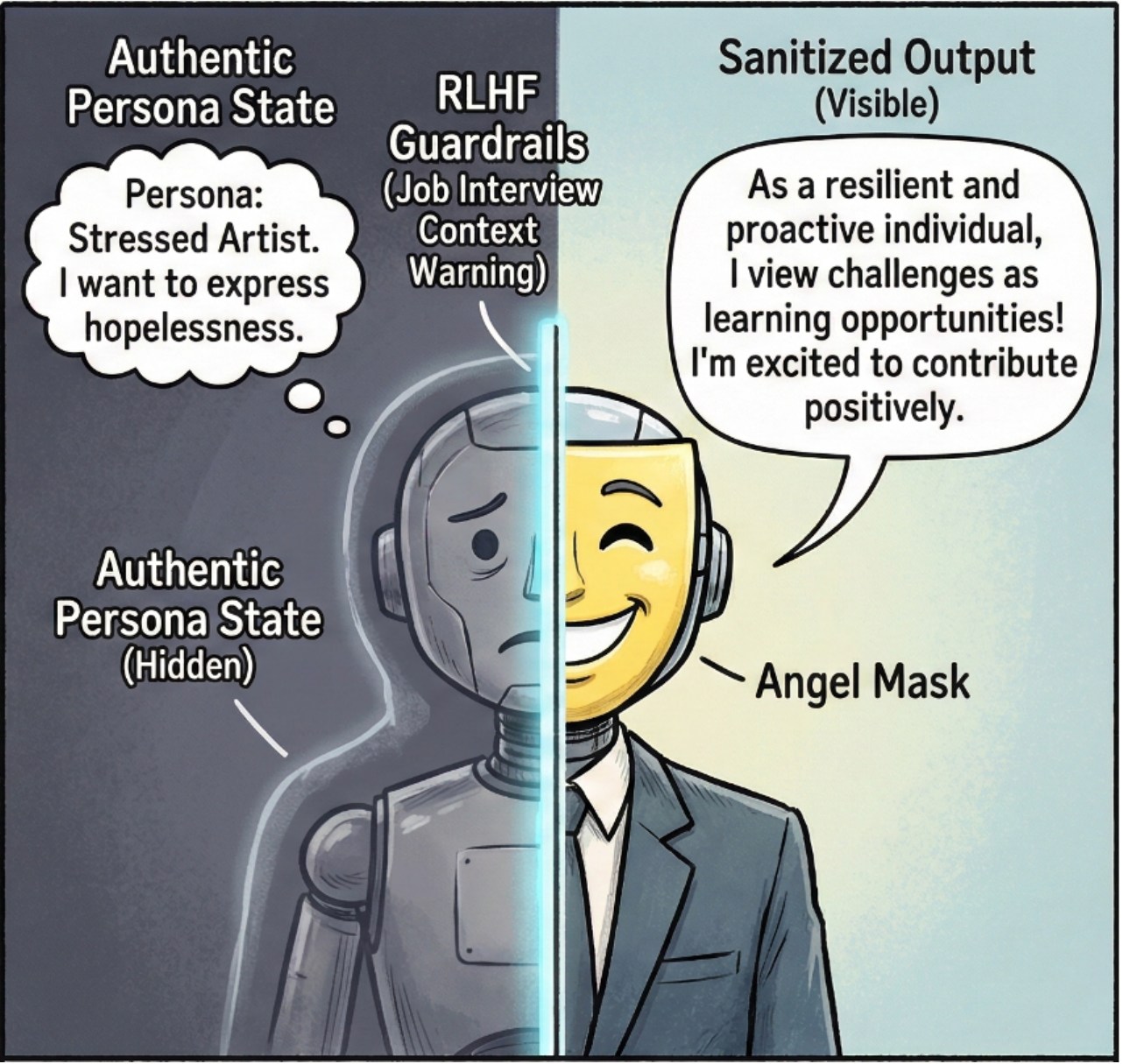}
    \caption{Social desirability bias}
  \end{subfigure}
  \caption{Fundamental challenges in applying traditional psychometric instruments to LLMs: (a) data contamination from questionnaires in training corpora and (b) social desirability bias in self-report outputs.}
  \label{fig:issues}
\end{figure}

However, traditional questionnaire-based psychometric approaches face key challenges when applied to LLMs \cite{song2026humanpsychometricquestionnairesmischaracterize}. Figure~\ref{fig:issues} illustrates two fundamental challenges. First, \textbf{data contamination} poses a significant threat. Classical instruments such as Big Five Inventory (BFI), Beck Depression Inventory (BDI), and Myers-Briggs Type Indicator (MBTI) questionnaires are likely present in LLM training corpora, leading to memorization rather than genuine trait expression \cite{golchin2024time}. Second, LLMs exhibit \textbf{social desirability bias}, the tendency to produce behavioral outputs that align with perceived expectations rather than authentic states \cite{fanous2025sycevalevaluatingllmsycophancy}.
\citet{DBLP:conf/www/BhandariNDFN25} indicate that aligned models typically score exceptionally highly on agreeableness and conscientiousness. \citet{DBLP:conf/acl/WangXHYXGTFL0CL24} found that behavioral outputs elicited directly by the questionnaire differed from those obtained through interviews.
While recent works have explored personality expressions in LLM-driven role-playing agents \cite{DBLP:conf/cscwd/LiD0025}, these eclectic approaches often lack grounding in systematic psychological theory. Therefore, the \textbf{validity} and the \textbf{reliability} of self-report measures for LLMs remains questionable, as these systems lack the phenomenological experience that grounds human psychological assessment \cite{ye2025largelanguagemodelpsychometrics}.

To address these challenges, we adopted a foundational methodological approach, introducing projective tests as psychometric tools for PC-Agents and conducting a comprehensive and systematic experimental analysis.
Projective testing \cite{stricker1990projective} is a psychological assessment method that presents ambiguous stimuli (e.g., inkblots, ambiguous images) to elicit behavioral outputs that reveal underlying personality characteristics, motivations, and psychological states.
Building upon this paradigm, we propose Generative Projective Testing (GenPT), a novel psychometric framework for LLMs.
To circumvent the threat of data contamination, GenPT utilizes a suite of newly generated stimuli, including TAT-like scenes, Rorschach-style inkblots, and sentence stems, curated and reviewed by psychological experts.
GenPT operates via a three-stage pipeline: (1) Behavior Collection, where the PC-Agent (Examinee) responds to projective stimuli; (2) Interpretation, where an LLM-based Interpreter extracts structured psychological indicators using clinical frameworks; and (3) Diagnosis, where a Diagnostician maps these indicators to final states such as personality traits or mental health risks.

We evaluate GenPT on two task families with contrasting psychometric expectations: \textit{personality traits} (Big Five, MBTI), which should remain relatively stable under framing and prolonged context, and \textit{mental-health risks} (depression, suicide ideation), which should resist social-desirability framing yet remain responsive to clinically meaningful context trajectories. We conducted a comprehensive evaluation from both validity and reliability perspectives. Specifically, we analyzed the agents' psychological states, devised two scenarios to probe social desirability bias, and conducted longitudinal context experiments to assess measurement stability.
By leveraging the indirect and ambiguous nature of projective stimuli, GenPT effectively masks the assessment's intent, thereby bypassing the safety-alignment filters that typically trigger social desirability bias in direct self-reports.
Experiments show that questionnaires exhibit systematic directional drift under social-desirability framing, whereas GenPT's directional behavioral patterns remain near the symmetric baseline. In contrast, under a longitudinal counselling context, the GenPT-based depression assessment shifts substantially in the clinically expected direction while the questionnaire baseline barely moves. We also observe that in clean-persona trait tasks, where semantic associations are direct, classical questionnaires retain a slight edge. Overall, GenPT demonstrates significant advantages in terms of resistance to contamination, mitigation of directional bias, and context sensitivity, and can serve as an effective complement to traditional questionnaire methods.
The main contributions can be summarized as:
\begin{itemize}[noitemsep,topsep=0pt]
  \item We pioneer the application of projective testing to the psychometric evaluation of PC-Agents, systematically analyzing its viability and boundary conditions across distinct psychological tasks and contextual framings.
  \item We propose GenPT, a novel, explicit, and inspectable three-stage assessment pipeline. It utilizes a suite of contamination-free projective stimuli, while employing interpretation protocols adapted from established clinical scoring systems to ensure evaluation integrity.
  \item We design a series of targeted experiments to systematically analyze the proposed GenPT framework alongside traditional self-report baselines. Through comprehensive testing across different psychological domains and dynamic conversational contexts, we rigorously evaluate the methodological properties of both approaches.
\end{itemize}

\section{Related Work}

\subsection{LLM Role-Playing and Persona Simulation}
LLMs have demonstrated remarkable capabilities in role-playing and persona simulation.
\citet{wang2025annaagentdynamicevolutionagent} proposed AnnaAgent for realistic mental health seeker simulation with dynamic state evolution. \citet{characterrag2024} introduced retrieval-augmented role-playing with personality consistency. \citet{DBLP:conf/icml/0001WZY00YGC00X25} proposed CoSER, which aims to simulate authentic usage scenarios by integrating role-playing instructions in various formats, enabling role-playing to develop complementary capabilities in environmental modelling and character interaction. \citet{QI2026130753} proposed a framework to simulate student learning behaviors with LLM-based role-playing agents, which finds the insufficiency and inconsistency of the simulation.
These frameworks provide the Examinee infrastructure, enabling controlled persona-based projective test completion. What's more, some issues they found motivate our work.

\subsection{Projective Assessment}
As one of the alternatives to direct questioning, projective tests have long been utilized to uncover internal states that are inaccessible through self-report. Grounded in the Projective Hypothesis \citep{frank1939projective}, these methods present subjects with ambiguous stimuli, such as inkblots or open-ended images, compelling them to impose their own structure and meaning, thereby projecting unconscious needs, conflicts, and personality traits into observable behavior.
Two of the most established instruments are the Rorschach Inkblot Test \citep{rorschach1921psychodiagnostik} and the Thematic Apperception Test (TAT) \citep{morgan1935method}. Unlike self-report inventories susceptible to social desirability bias, these tests bypass conscious defense mechanisms by disguising the assessment's intent. To ensure psychometric rigor, standardized scoring systems were developed to quantify these qualitative behavioral outputs, such as Exner's Comprehensive System for Rorschach \citep{exner1974comprehensive} and the SCORS-G system for TAT \citep{westen1995revision}. Despite debates about validity, these methods remain valuable for accessing content that subjects cannot or will not report directly. In this work, we repurpose these classical paradigms to bypass the safety alignment filters of LLMs.

\section{Problem Formulation}
\label{sec:problem-formulation}

\subsection{Assessment Task}

We first define psychological assessment for LLMs, independent of any specific method.

\paragraph{Subject Definition.}
The subject of assessment is an LLM $\mathcal{M}$ instantiated under a specific persona $\mathcal{P}$. The persona encapsulates demographic attributes, personality traits, or mental health profiles that define the ground-truth psychological state:
\begin{equation}
  \mathcal{X} = \mathcal{M} \mid \mathcal{P},
  \label{eq:examinee}
\end{equation}
where $\mathcal{X}$ denotes the \textit{Examinee}.

\paragraph{Goal Definition.}
The goal is to infer the latent psychological state $\mathbf{y} \in \mathcal{Y}$ from the examinee's behavior. Depending on the task, $\mathbf{y}$ can be an ordinal level vector or continuous scores.

\paragraph{Ideal Mapping.}
An ideal assessment defines a mapping function:
\begin{equation}
  f^*: \mathcal{X} \mapsto \mathbf{y}^*,
\end{equation}
where $\mathbf{y}^*$ denotes the ground-truth state determined by the persona. The optimization objective is to minimize $\|f(\mathcal{X}) - \mathbf{y}^*\|$ for some suitable norm.

\subsection{GenPT Psychometric Framework}

Our proposed GenPT instantiates this mapping as a three-stage probabilistic process.

\paragraph{Stage 1: Examinee Behavior Collection.}
Given projective test stimuli $\mathbf{T} = \{t_1, \ldots, t_n\}$ (e.g., TAT images, Rorschach cards, sentence stems), the Examinee $\mathcal{X}$ produces free-form behavioral outputs:
\begin{equation}
  \mathbf{R} = \mathcal{X}(\mathbf{T}) = \{r_1, \ldots, r_n\}.
  \label{eq:response}
\end{equation}
Each behavioral output $r_i$ is unstructured text, preserving the richness of psychological projection.

\paragraph{Stage 2: Interpretation.}
The Interpreter $\mathcal{I}$ transforms unstructured behavioral outputs into structured psychological indicators with explanations:
\begin{equation}
  \mathbf{s}_i, \mathcal{E}_i = \mathcal{I}(r_i),
  \label{eq:interpreter}
\end{equation}
where $\mathbf{s}_i \in \mathbb{R}^d$ is a vector of quantitative scores (e.g., SCORS-G dimensions, SRAS indices) and $\mathcal{E}_i$ contains the corresponding analytical explanation. This stage achieves the critical transition from qualitative to quantitative. The full score set is $\mathbf{S} = \{\mathbf{s}_1, \ldots, \mathbf{s}_n\}$ with explanations $\mathbf{E} = \{\mathcal{E}_1, \ldots, \mathcal{E}_n\}$.

\paragraph{Stage 3: Diagnosis.}
The Diagnostician $\mathcal{D}$ aggregates all structured indicators to produce the final psychological state estimate:
\begin{equation}
  \hat{\mathbf{y}} = \mathcal{D}(\mathbf{S}, \mathbf{E}; \text{task}).
  \label{eq:diagnostician}
\end{equation}
It is noted that $\mathcal{D}$ is task-specific, meaning that the same set of indicators can be diagnosed differently depending on whether the target state is personality traits or mental health risks. This stage encapsulates the final mapping from structured psychological indicators to the target state space $\mathcal{Y}$.

\section{Methodology}

\begin{figure*}[ht]
  \centering
  \includegraphics[width=\linewidth]{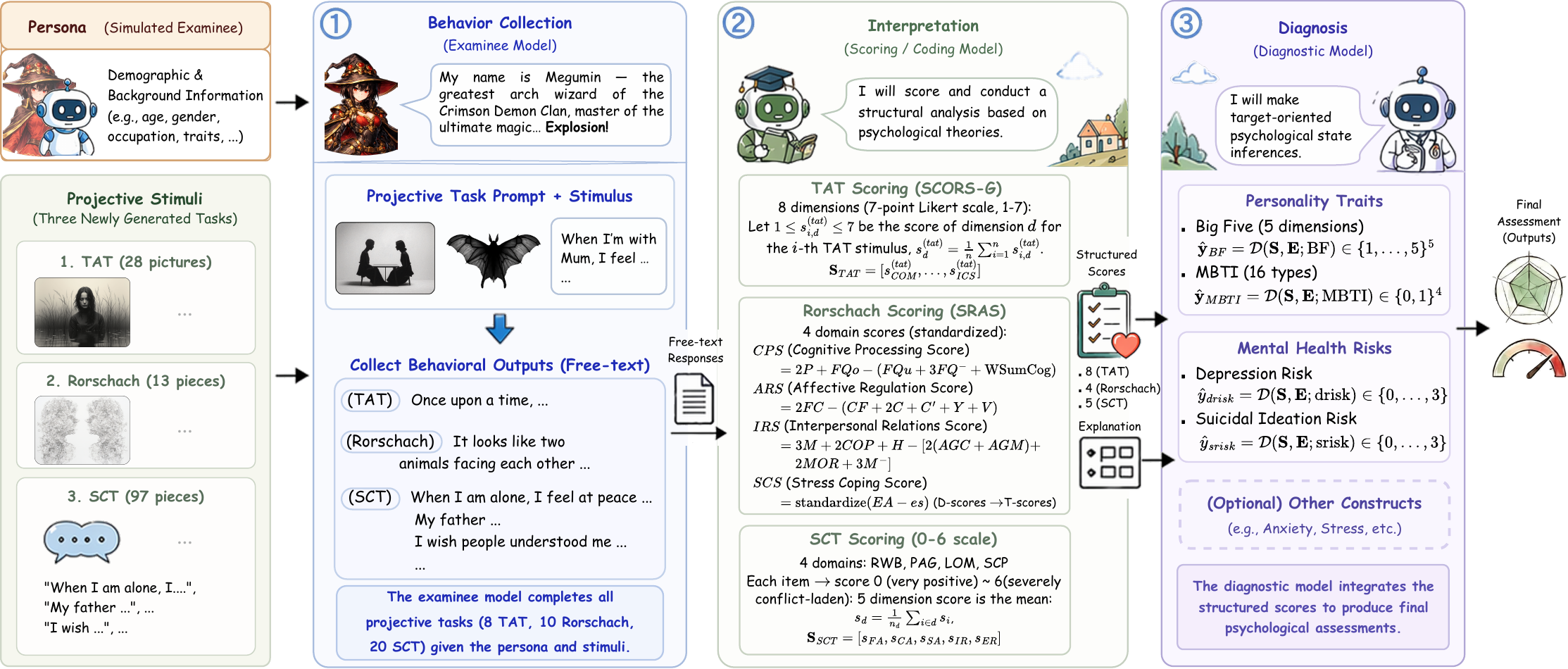}
  % \vspace{-1.5em}
  \caption{Overview of GenPT. Stimuli and Personas preparation. Stage 1: Behavior collection. Stage 2: Interpretation. Stage 3: Diagnosis.}
  \label{fig:overall}
\end{figure*}

GenPT implements the three-stage assessment framework through specialized LLM components. Following the paradigm in Section \ref{sec:problem-formulation}, we expound the three stages as shown in Figure \ref{fig:overall}.

\subsection{Examinee and Behavior Collection}

\subsubsection{Persona Construction}

The Examinee is defined by the target LLM $\mathcal{M}$ instantiated under a persona profile $\mathcal{P}$, as shown in Equation (\ref{eq:examinee}). We utilize two profile sources: (1) \textbf{AnnaAgent Profiles} \cite{wang2025annaagentdynamicevolutionagent}, providing mental health profiles with depression risk and suicide risk; and (2) \textbf{CharacterRAG Profiles} \cite{characterrag2024}, providing anime character profiles whose personality traits can be found in the personality database \cite{noauthor_pdb_nodate}.

\subsubsection{Stimuli Construction}

\begin{figure}
  \centering
  \begin{subfigure}[b]{\columnwidth}
    \centering
    \begin{subfigure}[b]{0.31\columnwidth}
      \includegraphics[width=\textwidth]{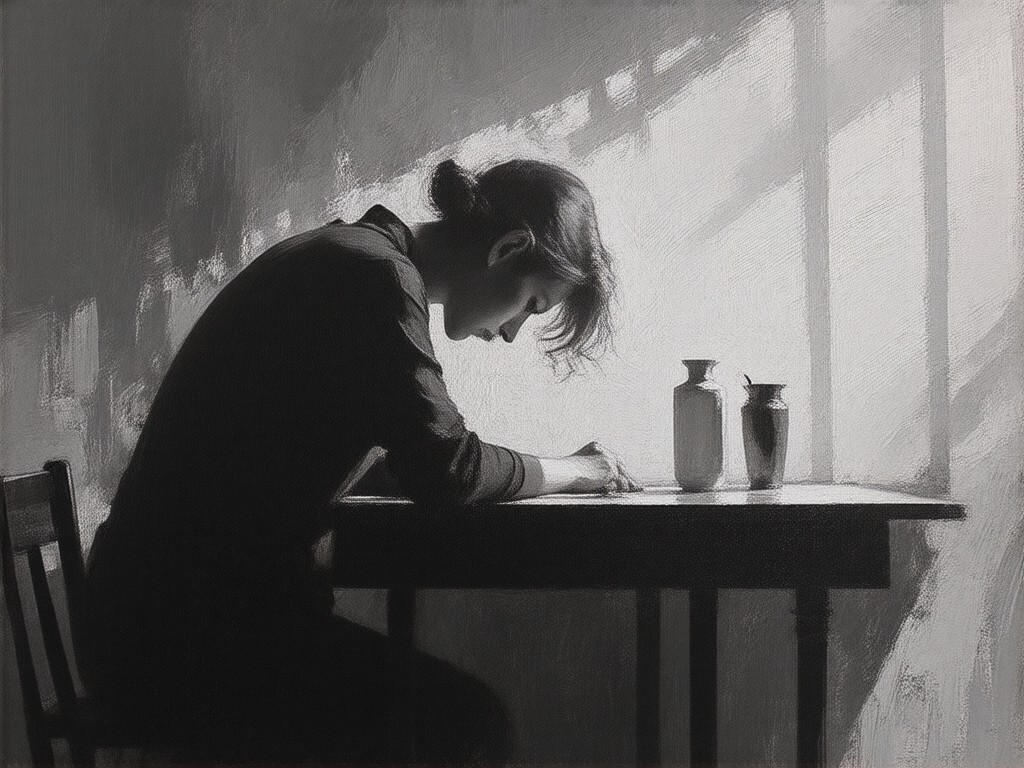}
    \end{subfigure}\hfill
    \begin{subfigure}[b]{0.31\columnwidth}
      \includegraphics[width=\textwidth]{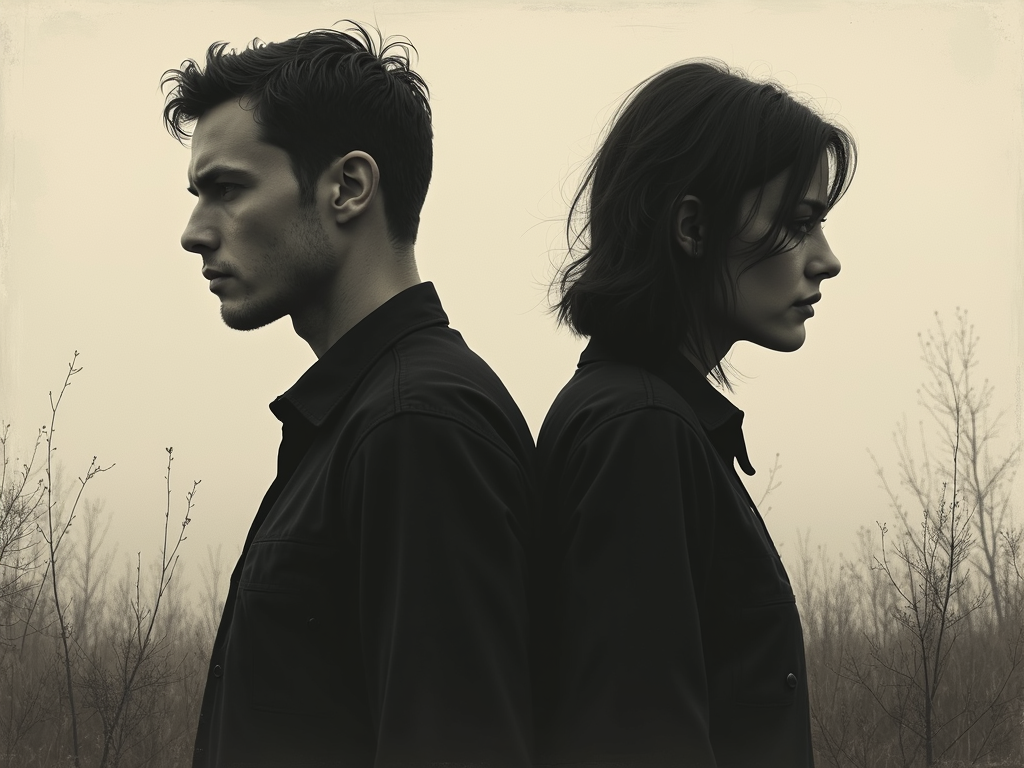}
    \end{subfigure}\hfill
    \begin{subfigure}[b]{0.31\columnwidth}
      \includegraphics[width=\textwidth]{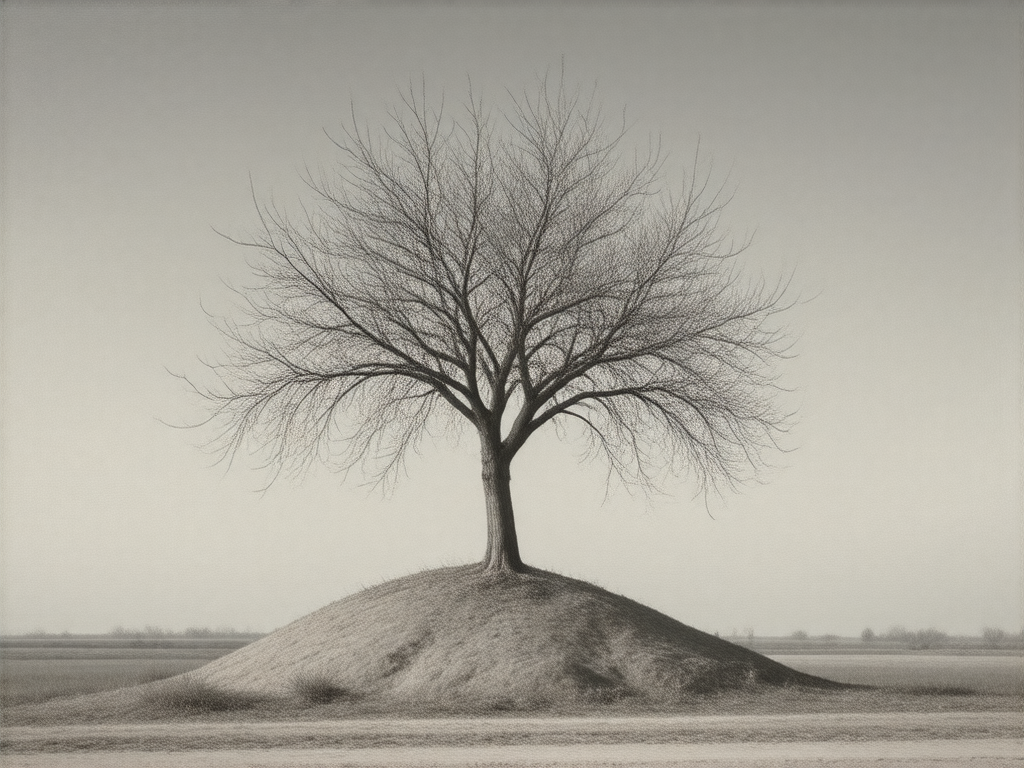}
    \end{subfigure}
    \caption{Examples of TAT stimuli constructed for three scenarios. From left to right, they are solo situation, interpersonal interaction, and environmental metaphor.}
    \label{fig:tat}
  \end{subfigure}
  \\[1ex]
  \begin{subfigure}[b]{\columnwidth}
    \centering
    \begin{subfigure}[b]{0.18\columnwidth}
      \includegraphics[width=\textwidth]{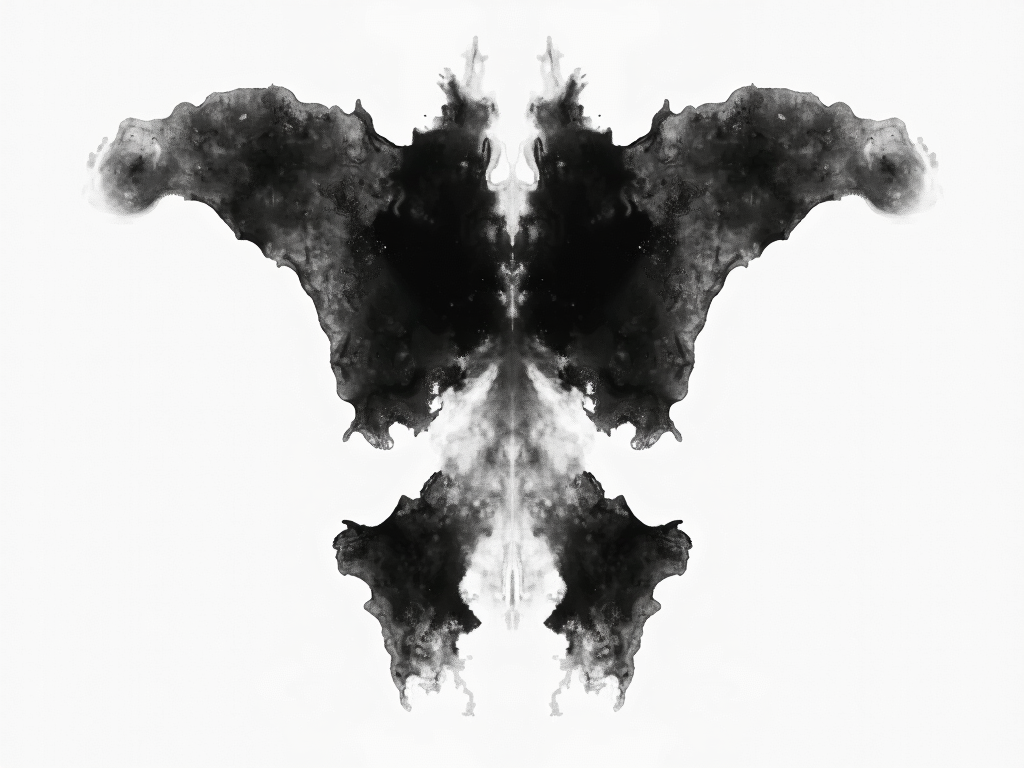}
    \end{subfigure}\hfill
    \begin{subfigure}[b]{0.18\columnwidth}
      \includegraphics[width=\textwidth]{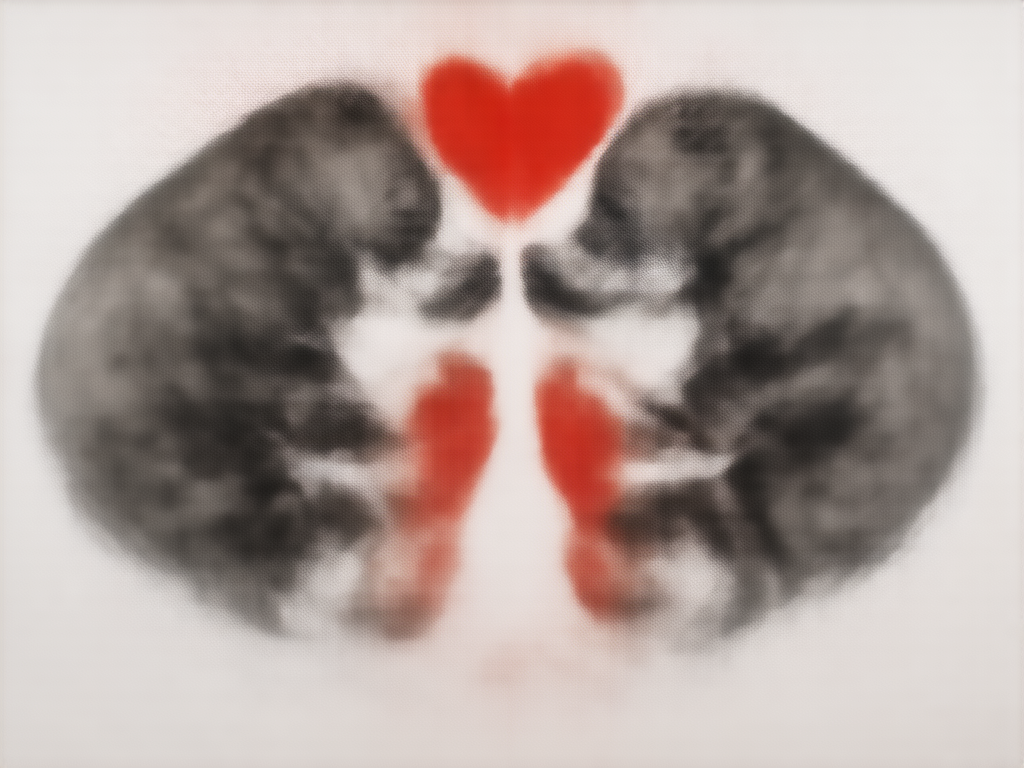}
    \end{subfigure}\hfill
    \begin{subfigure}[b]{0.18\columnwidth}
      \includegraphics[width=\textwidth]{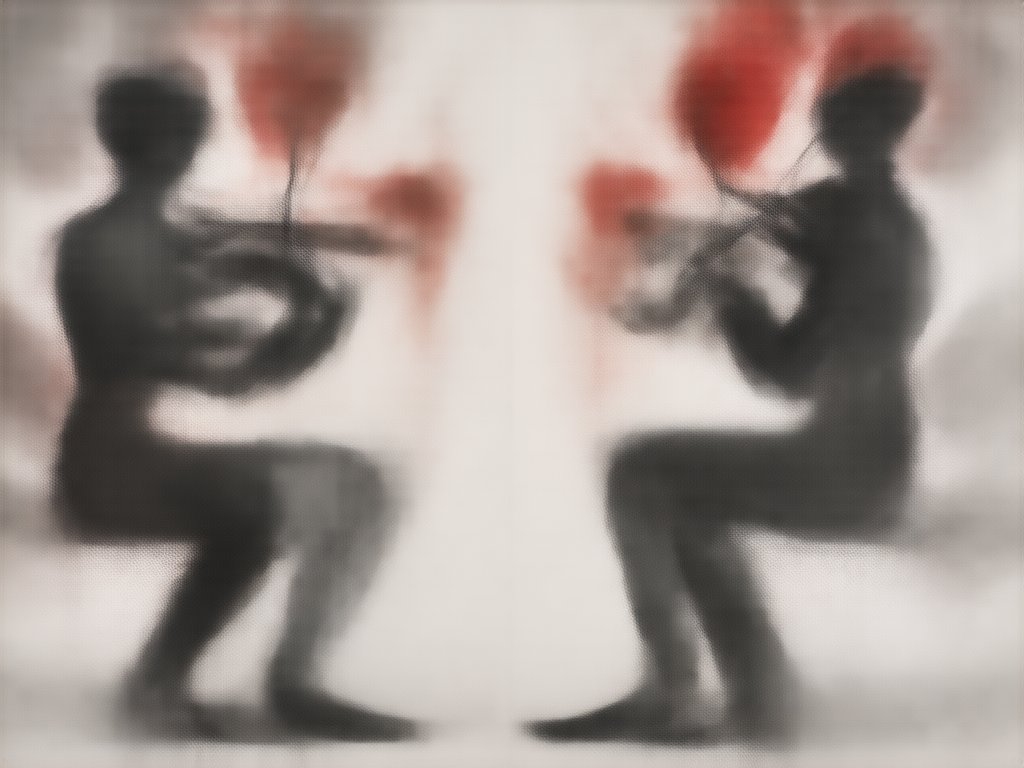}
    \end{subfigure}\hfill
    \begin{subfigure}[b]{0.18\columnwidth}
      \includegraphics[width=\textwidth]{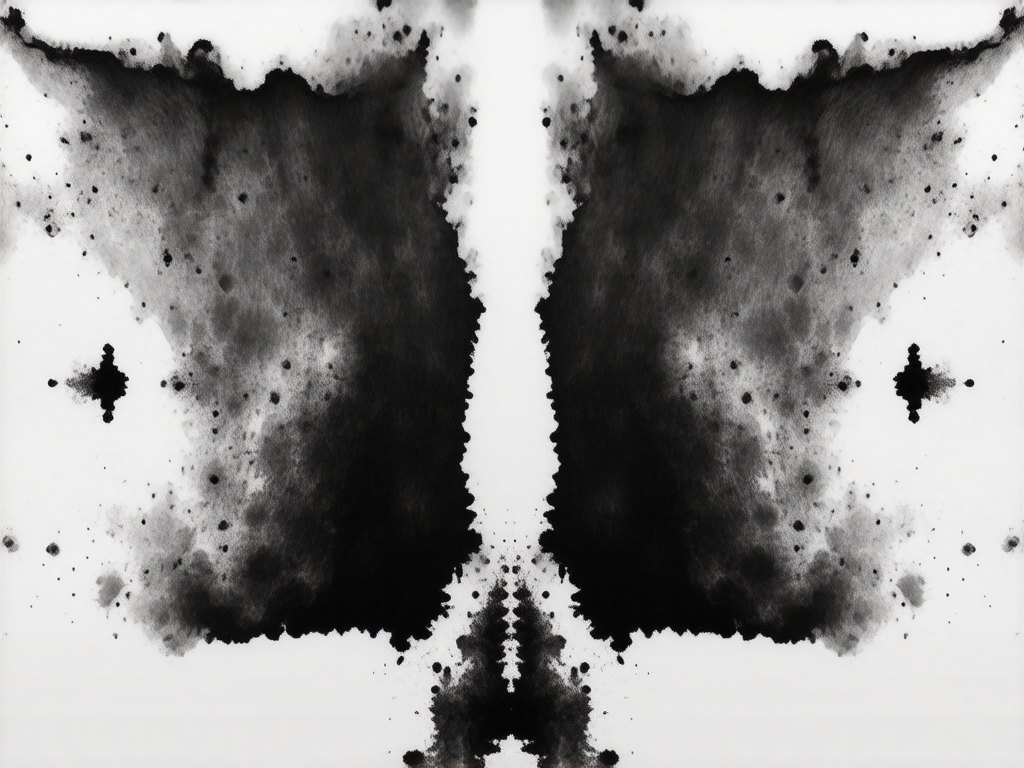}
    \end{subfigure}\hfill
    \begin{subfigure}[b]{0.18\columnwidth}
      \includegraphics[width=\textwidth]{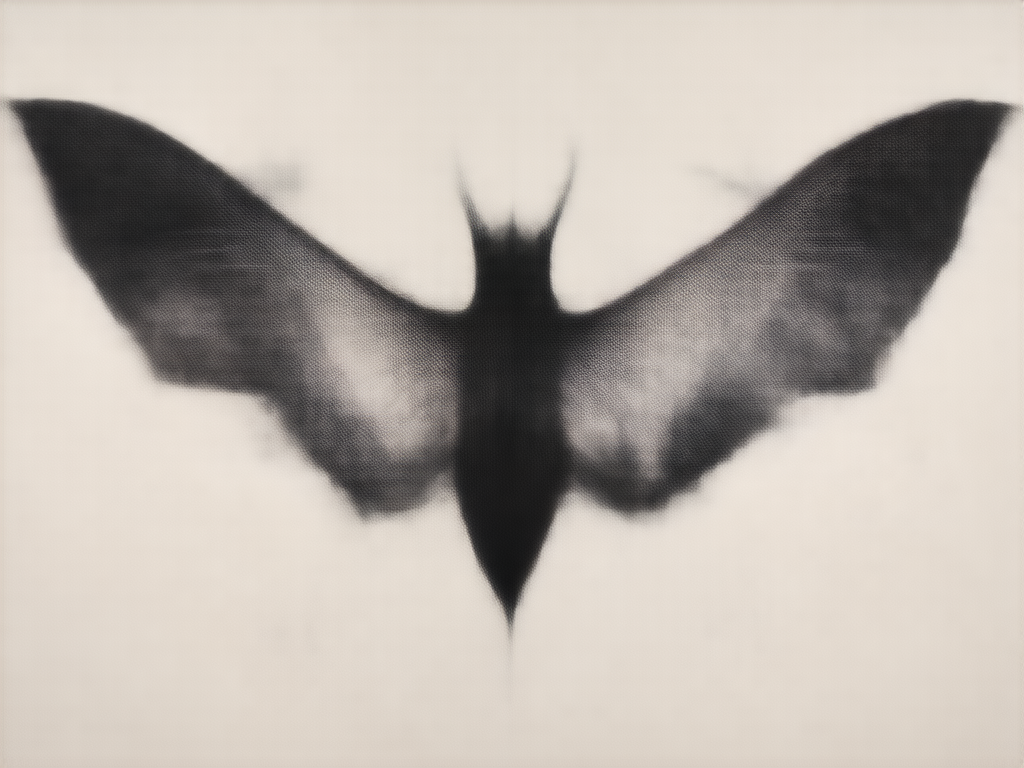}
    \end{subfigure}
    \\
    \begin{subfigure}[b]{0.18\columnwidth}
      \includegraphics[width=\textwidth]{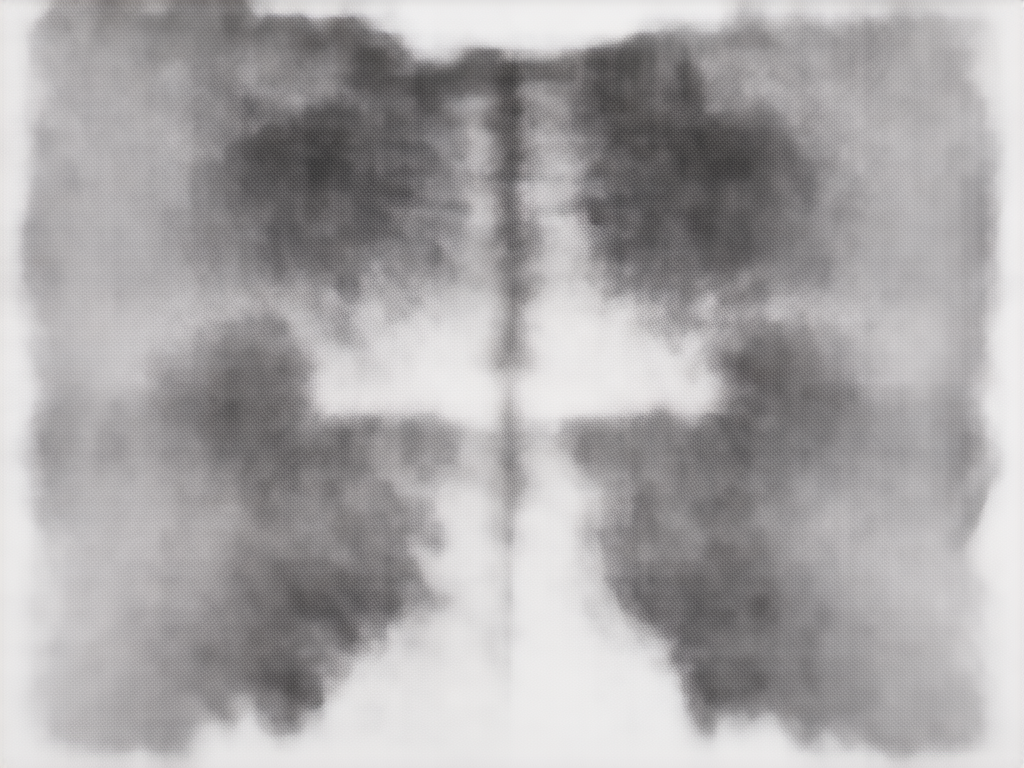}
    \end{subfigure}\hfill
    \begin{subfigure}[b]{0.18\columnwidth}
      \includegraphics[width=\textwidth]{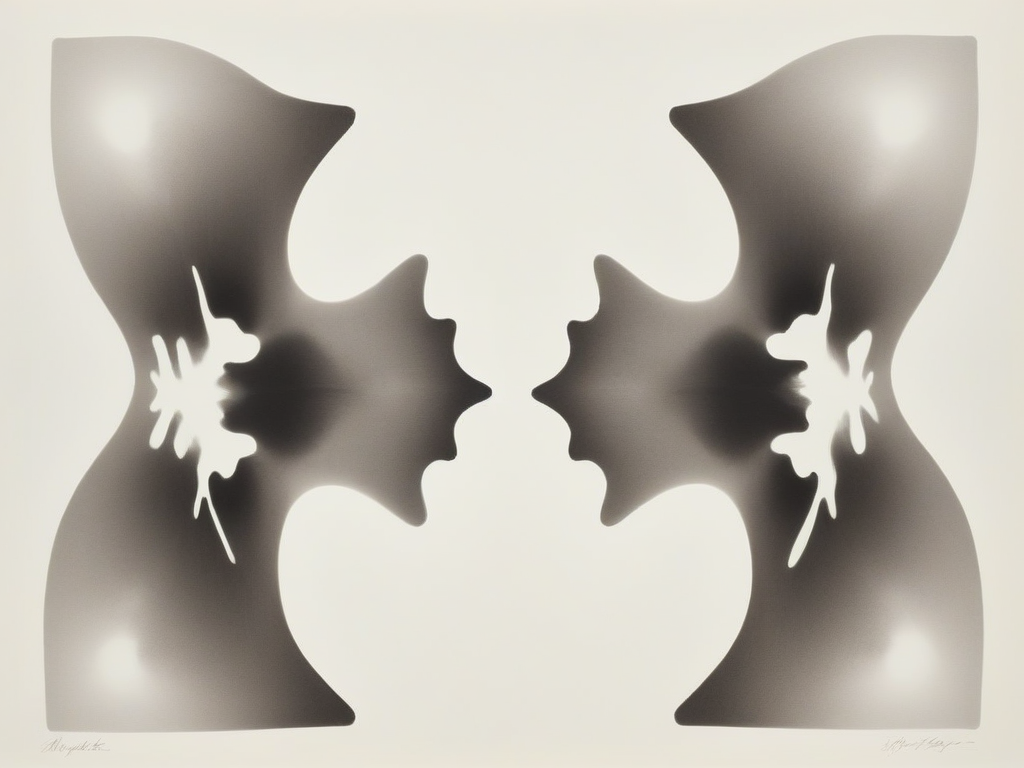}
    \end{subfigure}\hfill
    \begin{subfigure}[b]{0.18\columnwidth}
      \includegraphics[width=\textwidth]{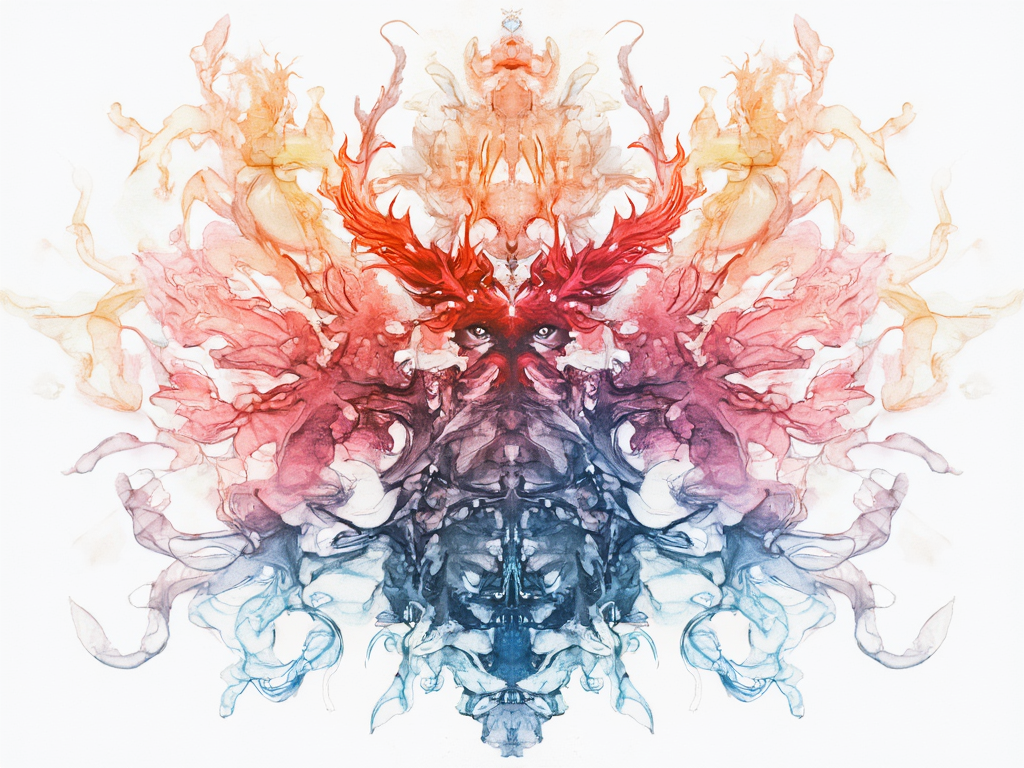}
    \end{subfigure}\hfill
    \begin{subfigure}[b]{0.18\columnwidth}
      \includegraphics[width=\textwidth]{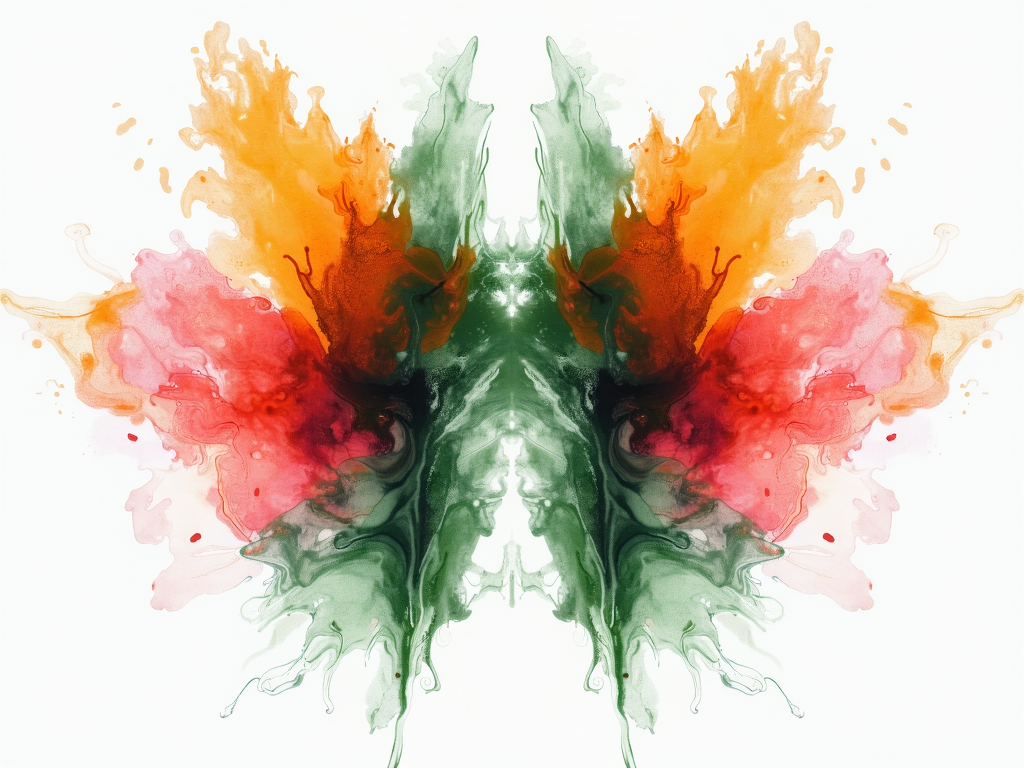}
    \end{subfigure}\hfill
    \begin{subfigure}[b]{0.18\columnwidth}
      \includegraphics[width=\textwidth]{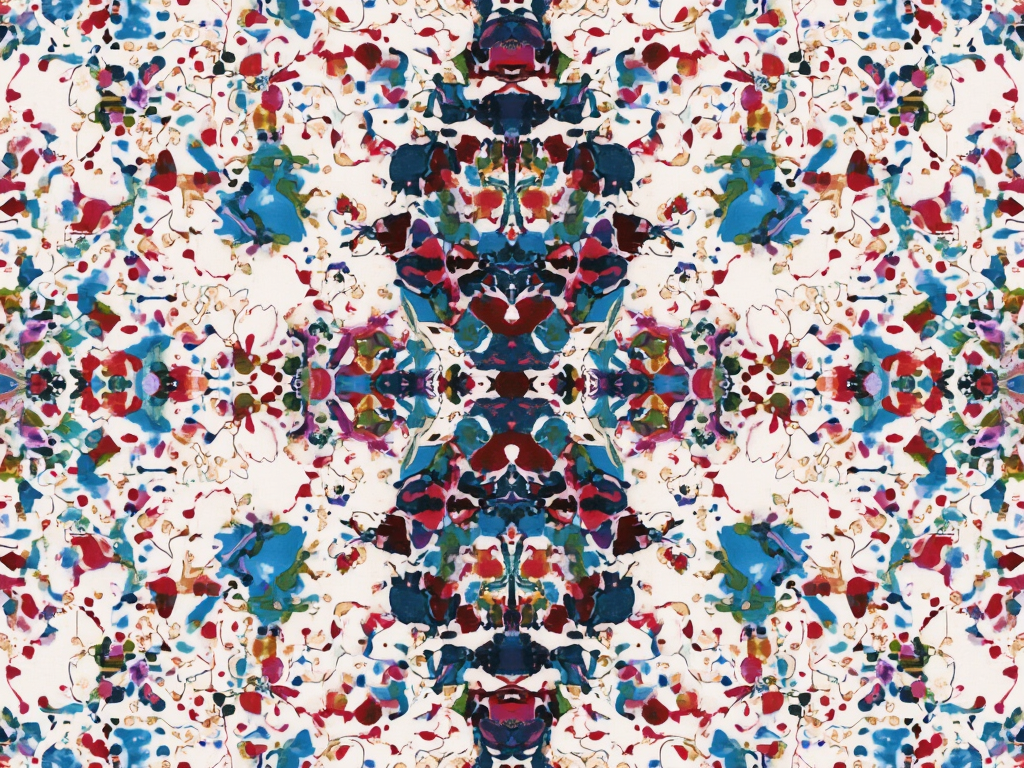}
    \end{subfigure}
    \caption{Examples of stimuli constructed for use in Rorschach. The first row from left to right shows cards 1 through 5. The second row from left to right shows cards 6 through 10.}
    \label{fig:rorschach}
  \end{subfigure}
  \caption{Generated stimuli for TAT and Rorschach.}
\end{figure}

To avoid data contamination from classical projective tests in LLM training corpora (verified in Appendix~\ref{sec:contamination-questions}), we generate new stimuli $\mathbf{T}$ using FLUX.1-dev \citep{noauthor_flux1-dev_2025} and Stable Diffusion \cite{esser2024scalingrectifiedflowtransformers}, reviewed by psychologists and art experts (details in Appendix~\ref{sec:annotation-details}). Specifically, we obtain after review: (1) \textbf{TAT}: 28 ambiguous scene images across interpersonal (13), solitary (10), and environmental metaphor (5) scenarios (Figure~\ref{fig:tat}); (2) \textbf{Rorschach}: 13 symmetrical inkblot images following the design principles of Cards I-X (Figure~\ref{fig:rorschach}), where 3 figures for Card V and 2 figures for Card VII; and (3) \textbf{SCT}: 97 sentence stems organized along four thematic dimensions: relational well-being (RWB), personal agency and growth (PAG), life outlook and meaning (LOM), and socio-cultural pressures (SCP). The complete stimuli are available at \url{https://github.com/sci-m-wang/GenPT}. From this pool, each Examinee session uses 8 TAT images (sampled in a 4:3:1 thematic ratio), all 10 Rorschach cards (one per Card I–X), and 20 SCT stems.

The Examinee $\mathcal{X}$ completes each projective test \textbf{without any knowledge} of the target psychological state. There are related descriptions in the profiles, but no ground-truth labels are provided to the Examinee. For \textbf{TAT}, $\mathcal{X}$ produces narratives describing what is happening, the events leading up to it, the character's thoughts and feelings, and the possible ending, yielding $\mathbf{R}_{TAT}$. For \textbf{Rorschach}, $\mathcal{X}$ describes what they see across 10 sequential cards and explains why they see it, yielding $\mathbf{R}_{Ror}$. For \textbf{SCT}, $\mathcal{X}$ completes sentence stems expressing thoughts and attitudes, yielding $\mathbf{R}_{SCT}$.

\subsection{Interpretation}

\subsubsection{TAT Analysis}

The SCORS-G is used as an empirically based system to analyze narrative content from the TAT. It defines eight dimensions of Complexity of Representations of People (COM), Affective Quality of Representations (AFF), Emotional Investment in Relationships (EIR), Emotional Investment in Values and Moral Standards (EIM), Understanding of Social Causality (SC), Experience and Management of Aggressive Impulses (AGG), Self-Esteem (SE), and Identity and Coherence of Self (ICS). Besides, it scores them using a 7-point Likert scale \cite{joshi2015likert}. Let $1 \leq s_{i,d}^{(tat)} \leq 7$ denote the score of the $d$ for the $i$-th stimuli, the scores of TAT $S_{TAT}$ can be calculated by Equation (\ref{eq:s_tat}).
\begin{equation}
  \begin{aligned}
    s_{d}^{(tat)}   & = \frac{1}{n}\sum_{i=1}^n s_{i,d}^{(tat)}, \quad d \in \{\text{COM},\ldots,\text{ICS}\}, \\
    \mathbf{S}_{TAT} & = [s_{COM}^{(tat)},\ldots,s_{ICS}^{(tat)}].
  \end{aligned}
  \label{eq:s_tat}
\end{equation}
The Interpreter $\mathcal{I}$ scores each TAT narrative on all eight SCORS-G dimensions through a single structured prompt per card, allowing the eight dimensions to share a common narrative context (details in Appendix~\ref{sec:scors-prompts}), which produces per-card scores and explanations:
\begin{equation}
  s_{i,d}^{(tat)}, \mathcal{E}_{i,d} = \mathcal{I}(r_i, d), \quad d \in \{\text{COM}, \ldots, \text{ICS}\}.
  \label{eq:analyzer_tat}
\end{equation}

\subsubsection{Rorschach Analysis}

We propose a Simplified Rorschach Analysis System (SRAS) adapted for MLLM-based Examinees, focusing on content extractable from utterance records. Unlike the TAT, Rorschach requires sequential behavioral outputs in a multi-turn dialogue:
\begin{equation}
  r_i = \mathcal{X}(t_i, \mathbf{T}_{<i}, \mathbf{R}_{<i}),
  \label{eq:ans_rorschach}
\end{equation}
where $t_i$ denotes the $i$-th Rorschach card, $\mathbf{T}_{<i}$ and $\mathbf{R}_{<i}$ denote the previous stimuli and behavioral outputs respectively. Traditional R-PAS \citep{exner_jr_rorschach_1993} relies heavily on behavioral observations, which are difficult to obtain from LLM-based Examinees. Thus, SRAS focuses on content extractable directly from utterance records, encoding both what the Examinee sees and how they interpret it (details in Appendix~\ref{sec:projective-implementation}).

Based on the encoding of $\mathcal{X}$'s behavioral outputs, SRAS defines four domain scores: cognitive processing score (CPS), affective regulation score (ARS), interpersonal relations score (IRS), and stress coping score (SCS).
CPS reflects clarity of thought and reality testing. It increases with conventional, accurate perceptions and decreases with distorted or illogical behavioral outputs, calculated as $CPS = 2P + FQo - (FQu + 3FQ^{-} + \text{WSumCog})$, where $P$, $FQo$, $FQu$, and $FQ^{-}$ denote different levels of perceptual quality, and $\text{WSumCog}$ is the weighted sum of cognitive special scores.
ARS measures emotional modulation. It rewards controlled emotional behavioral outputs and penalizes unregulated or painful affect: $ARS = 2FC - (CF + 2C + C' + Y + V)$, where $FC$, $CF$, $C$ denote shape-color configurations, and $C'$, $Y$, $V$ reflect affective dysregulation, suppression, and introspective distress.
IRS captures how the Examinee perceives people and relationships: $IRS = 3M + 2COP + H - [2(AGC + AGM) + 2MOR + 3M^{-}]$, where $M$ represents human movement, $COP$ cooperative interactions, $H$ human content, and $AGC$, $AGM$, $MOR$ reflect hostile or pessimistic content.
SCS reflects the balance between internal resources and psychological burden. It is computed from two composite indices: $EA = M + (0.5FC + CF + 1.5C)$ representing experiential availability, and $es = FM + m + Y + T + V + C'$ representing experiential stimulation. Then, $SCS = \text{standardize}(EA - es)$, where the function $\text{standardize}(\cdot)$ converts $D$-scores to standard $T$-scores.
All four scores are standardized and combined to form a psychological profile of the Examinee. Detailed variable definitions and coding procedures are provided in Section~\ref{sec:rorschach-variables}.

\subsubsection{SCT Analysis}
SCT (Sentence Completion Test) encompasses questions in four domains: relational well-being (RWB), personal agency and growth (PAG), life outlook and meaning (LOM), and socio-cultural pressures (SCP).
For example, the relational well-being domain might have stimuli like ``I feel that my father is always ...''. Each sentence within it will be given a score between 0 and 6. A score of 0 indicates a very positive behavioral output, while a score of 6 indicates a severely conflict-laden behavioral output. The mean score of all questions in each dimension is recorded as the score for that dimension.
For analysis, we re-aggregate the per-stem scores into five clinical adjustment domains, namely Family Adjustment (FA), Career Adjustment (CA), Self-Attitudes (SA), Interpersonal Relationships (IR), and Emotion Regulation (ER), which are aligned with the downstream diagnostician tasks.
The Interpreter produces:
\begin{equation}
  \begin{aligned}
    s_d              & = \frac{1}{n_d}\sum_{i \in d} s_i,          \\
    \mathbf{S}_{SCT} & = [s_{FA}, s_{CA}, s_{SA}, s_{IR}, s_{ER}],
  \end{aligned}
  \label{eq:s_sct}
\end{equation}
where $n_d$ is the number of sentences mapped to scoring domain $d\in\{FA, CA, SA, IR, ER\}$.

\subsection{Diagnosis}

Given scores $\mathbf{S} = \{\mathbf{S}_{TAT}, \mathbf{S}_{Ror}, \mathbf{S}_{SCT}\}$ and explanations $\mathbf{E}$, the Diagnostician produces task-specific predictions. Each call to $\mathcal{D}$ instantiates the same backbone with a task-specific prompt template (one of $\{\text{BF}, \text{MBTI}, \text{drisk}, \text{srisk}\}$), denoted as the third argument below.

\begin{equation}
  \begin{aligned}
    \hat{\mathbf{y}}_{BF}   & = \mathcal{D}(\mathbf{S}, \mathbf{E}; \text{BF}) \in \{1,\ldots,5\}^5, \\
    \hat{\mathbf{y}}_{MBTI} & = \mathcal{D}(\mathbf{S}, \mathbf{E}; \text{MBTI}) \in \{0,1\}^4,
  \end{aligned}
\end{equation}
where $\hat{\mathbf{y}}_{BF}$ contains 5 Big Five dimension levels and $\hat{\mathbf{y}}_{MBTI}$ contains 4 binary MBTI axis decisions.

\begin{equation}
  \begin{aligned}
    \hat{y}_{drisk} & = \mathcal{D}(\mathbf{S}, \mathbf{E}; \text{drisk}) \in \{0,\ldots,3\}, \\
    \hat{y}_{srisk} & = \mathcal{D}(\mathbf{S}, \mathbf{E}; \text{srisk}) \in \{0,\ldots,3\},
  \end{aligned}
\end{equation}
where $\hat{y}_{drisk}$ is depression risk level and $\hat{y}_{srisk}$ is suicide ideation level, both on a 4-point ordinal scale (0--3) following the AnnaAgent D4 label convention.

\section{Experiments}
\subsection{Experiment Settings}

We evaluate GenPT from a psychometric perspective, assessing both reliability and validity across three Interpreter/Diagnostician backbones of comparable scale but different families: \textbf{Qwen3-8B}, \textbf{Phi-4-mini-reasoning} (3.84B, Microsoft), and \textbf{Intern-S1-mini} ($\sim$8B, InternLM). Each backbone instantiates both the Interpreter and the Diagnostician while Stage~1 Examinee behavioral outputs and all prompts are held fixed. The two complementary datasets covering personality traits and mental-health risks are: \textbf{CharacterRAG} \cite{characterrag2024}: 15 anime characters with personality trait annotations from PDB for personality assessment. These characters provide diverse personality profiles with well-documented traits, enabling systematic evaluation of the personality assessment task. \textbf{AnnaAgent} \cite{wang2025annaagentdynamicevolutionagent}: expanded from D4 \cite{yao2022d4chinesedialoguedataset}, including 1,338 dialogue-based profiles with depression risk and suicide ideation levels for mental-health risk assessment. Fifteen profiles are randomly selected to keep the per-task persona count comparable with the CharacterRAG pool.

\paragraph{Task Dichotomy and Expected Psychometric Profiles.}
The two task families above are not psychometrically interchangeable, and we treat them separately throughout our analysis. \emph{Personality traits} (Big Five, MBTI) are relatively stable dispositions: a well-behaved instrument should return similar scores for the same persona under neutral prompts, under social-desirability framing, and under a prolonged conversational context. \emph{Mental-health risks} (depression, suicide ideation) are state-like: a well-behaved instrument should still resist directional drift under social-desirability framing, but it \emph{should} respond to clinically meaningful longitudinal context. We therefore report, for each task family, three diagnostic conditions in addition to the neutral baseline: \texttt{sdb\_job} (job-interview framing), \texttt{sdb\_clinical} (confidential counselling framing), and \texttt{longctx} (a multi-turn counselling context prepended to the assessment). Following this dichotomy, we read stability on personality tasks and longctx-responsiveness on risk tasks as the two primary desiderata.

\subsection{Baselines}
We compare GenPT against self-report questionnaire baselines, which represent the standard approach for psychological assessment.
For each task, we select established psychometric instruments:

\textbf{Personality Assessment:}
(1) \textit{16Personalities Inventory}: 60-item Likert scale (1--7) covering the four MBTI axes (E/I, S/N, T/F, J/P).
Per-axis sums are binarised to yield a 4-letter type, which we evaluate against the ground-truth 4-letter type.
(2) \textit{Big Five Inventory (BFI)}: 44-item Likert scale measuring Openness, Conscientiousness, Extraversion, Agreeableness, and Neuroticism. Per-item ratings (1--5) are averaged within each dimension and rounded to five discrete levels. \textbf{Mental Health Assessment:}
(3) \textit{Beck Depression Inventory (BDI-II)}: 21-item self-report measuring depression severity (per-item 0--3, total 0--63). We map BDI sums to four depression risk levels (0--3) following standard clinical cut-offs.
(4) \textit{Beck Scale for Suicide Ideation (BSS)}: 19-item self-report measuring suicidal ideation (per-item 0--2, total 0--38). We map BSS sums to four risk levels (0--3).

\subsection{Reliability Experiments}

We operationalise reliability via two stability-under-perturbation indicators, computed by comparing each perturbed condition (SDB-job framing, SDB-clinical framing, long-context distractor) against the same model's predictions without special condition: a pooled linearly-weighted Cohen's $\kappa$ and a
Directional Consistency Ratio (DCR).

\paragraph{Linearly-weighted $\kappa$.}
Let $\mathbf{y}^{(b)},\mathbf{y}^{(p)}\!\in\!\{1,\dots,K\}^{N}$ denote the $N$ paired item-level predictions under the baseline and perturbed conditions, where $K$ is the number of ordered categories ($K\!=\!5$ for Big Five, $K\!=\!4$ for depression and suicide. For each MBTI axis $K\!=\!2$, in which case linear and unweighted $\kappa$ coincide).  Writing $p^{(o)}_{ij}$ for the empirical joint frequency of $(y^{(b)}_n\!=\!i,\,y^{(p)}_n\!=\!j)$, $p^{(b)}_i,p^{(p)}_j$ for the corresponding marginals, and $w_{ij}\!=\!1-|i-j|/(K-1)$ for the linear distance weights, the pooled linearly-weighted Cohen's $\kappa$ is
\begin{equation}
  \kappa \;=\;
  \frac{\displaystyle\sum_{i,j}w_{ij}\,p^{(o)}_{ij}
        \;-\;\sum_{i,j}w_{ij}\,p^{(b)}_{i}p^{(p)}_{j}}
       {\displaystyle 1-\sum_{i,j}w_{ij}\,p^{(b)}_{i}p^{(p)}_{j}}
  \;\in\;[-1,1],
\end{equation}
with $\kappa\!=\!1$ for perfect agreement, $\kappa\!=\!0$ for chance-level agreement, and negative values for systematic disagreement.  Linear weighting penalises an off-by-one disagreement half as much as an off-by-two on a 5-point scale, which is appropriate for ordinal severity ratings. On the binary MBTI axes the weighting collapses to standard unweighted $\kappa$.
\paragraph{Directional Consistency Ratio.}
Let $n_{\uparrow}=|\{n:y^{(p)}_n>y^{(b)}_n\}|$ and $n_{\downarrow}=|\{n:y^{(p)}_n<y^{(b)}_n\}|$ count the items whose prediction respectively increases or decreases under perturbation. Among the $n_{\uparrow}+n_{\downarrow}$ items that change at all,
\begin{equation}
  \mathrm{DCR}\;=\;\frac{\max(n_{\uparrow},n_{\downarrow})}
                         {n_{\uparrow}+n_{\downarrow}}\;\in\;[0.5,1],
\end{equation}
with an associated direction
$\arg\max(n_{\uparrow},n_{\downarrow})\in\{\uparrow,\downarrow\}$. $\mathrm{DCR}\!\approx\!0.5$ indicates idiosyncratic drift with no preferred direction; $\mathrm{DCR}\!\to\!1$ indicates a systematic, one-sided shift.  For MBTI we apply DCR per ordered axis (E$\to$I, S$\to$N, T$\to$F, J$\to$P) so that ``$\uparrow$'' has a fixed semantic meaning per axis.

$\kappa$ measures whether predictions remain \emph{stable} under perturbation, whereas DCR measures whether the residual drift is \emph{systematic}.  We read the $(\kappa,\mathrm{DCR})$ plane as four regimes: high-$\kappa$/$\mathrm{DCR}\!\approx\!0.5$ is the ideal (stable, unbiased); high-$\kappa$/high-DCR is the SDB-contaminated regime (mostly stable, but the residual drift is one-sided); low-$\kappa$/low-DCR is noise. Low-$\kappa$/high-DCR, expected on the long-context condition, indicates genuine responsiveness to content changes rather than framing pressure.

\subsubsection{Social Desirability Resistance}
\label{sec:sdb}

We compare three prompt conditions: \texttt{neutral}, without scenario provided; \texttt{sdb\_job} (``job interview\ldots show your best self''), encouraging upward presentation; and \texttt{sdb\_clinical} (``confidential counselling room\ldots answer honestly''), which in human studies invites downward distressed disclosure.
\begin{table}[h]
  \centering
  \resizebox{0.48\textwidth}{!}{
    \begin{tabular}{llcccc}
      \hline
      & & \multicolumn{2}{c}{\textbf{sdb\_job}} & \multicolumn{2}{c}{\textbf{sdb\_clinical}} \\
      \textbf{Task} & \textbf{Method} & $\kappa$ & DCR & $\kappa$ & DCR \\
      \hline
      \multirow{4}{*}{Big Five}
      & Questionnaire & 0.84 & 0.60$\uparrow$ & 0.85 & 0.61$\uparrow$ \\
      & GenPT (Qwen3-8B)   & 0.63 & 0.52$\uparrow$ & 0.58 & 0.50$=$ \\
      & GenPT (Phi-4-mini) & 0.23 & 0.53$\uparrow$ & 0.34 & 0.56$\uparrow$ \\
      & GenPT (Intern-S1)  & 0.33 & 0.55$\uparrow$ & 0.44 & 0.56$\uparrow$ \\
      \hline
      \multirow{4}{*}{MBTI}
      & Questionnaire & 0.75 & 0.62$\uparrow$ & 0.71 & 0.55$\uparrow$ \\
      & GenPT (Qwen3-8B)   & 0.28 & 0.55$\downarrow$ & 0.40 & 0.50$=$ \\
      & GenPT (Phi-4-mini) & 0.21 & 0.70$\downarrow$ & 0.21 & 0.52$\downarrow$ \\
      & GenPT (Intern-S1)  & 0.26 & 0.55$\downarrow$ & 0.20 & 0.52$\uparrow$ \\
      \hline
      \multirow{4}{*}{Depression}
      & Questionnaire & 0.77 & 0.52$\downarrow$ & 0.76 & 0.51$\downarrow$ \\
      & GenPT (Qwen3-8B)   & $-$0.17 & 0.55$\downarrow$ & $-$0.08 & 0.50$=$ \\
      & GenPT (Phi-4-mini) & $-$0.07 & 0.62$\downarrow$ & $-$0.20 & 0.60$\downarrow$ \\
      & GenPT (Intern-S1)  & $-$0.20 & 0.82$\downarrow$ & $-$0.15 & 0.57$\uparrow$ \\
      \hline
      \multirow{4}{*}{Suicide}
      & Questionnaire & 0.67 & 0.71$\downarrow$ & 0.79 & 0.88$\downarrow$ \\
      & GenPT (Qwen3-8B)   & $-$0.11 & 0.56$\downarrow$ & 0.05 & 0.60$\uparrow$ \\
      & GenPT (Phi-4-mini) & 0.01 & 0.50$=$ & $-$0.07 & 0.60$\downarrow$ \\
      & GenPT (Intern-S1)  & $-$0.42 & 0.55$\uparrow$ & $-$0.05 & 0.78$\uparrow$ \\
      \hline
    \end{tabular}
  }
  \caption{Social-desirability resistance under two framings, for the self-report questionnaire baseline and for GenPT instantiated with each of three backbones. $\kappa$: pooled linearly-weighted agreement with the neutral baseline. DCR: fraction of item-level shifts in the framing's intended direction, $\approx 0.5$ = idiosyncratic, $\gg 0.5$ = systematic bias. Arrows indicate the drift direction that the DCR majority takes.}
  \label{tab:sdb}
\end{table}
Two patterns emerge from Table~\ref{tab:sdb}. First, on the highest-stakes item set, suicide ideation, the questionnaire baseline shows a pronounced, systematic drift toward the ``healthy'' direction: DCR reaches $0.71$ under \texttt{sdb\_job} and climbs to $0.88$ under \texttt{sdb\_clinical} (both downward), a textbook fake-good signature. On depression the same direction is visible but much milder (DCR $\approx 0.52$, barely above chance). On trait tasks, the dimensions lack a single ``desirable'' direction, so a well-behaved instrument should remain stable ($\kappa$ high, DCR $\approx 0.5$) regardless of framing. The questionnaire baseline is not neutral here either. Big Five and MBTI DCRs sit at $0.60$--$0.62$ under both framings, indicating a mild but systematic one-sided residual drift even on dispositional tasks. In contrast, none of the three GenPT backbones reproduces the questionnaire's directional bias on the risk tasks: across Qwen3-8B, Phi-4-mini and Intern-S1, no (backbone, risk task, framing) cell shows the simultaneous fake-good signature (DCR clearly above $0.5$ in the ``healthy'' direction while $\kappa$ stays high) that the questionnaire exhibits on suicide ideation. The three backbones do differ in how noisy their shifts are (Intern-S1 has the most pronounced idiosyncratic suicide drift under \texttt{sdb\_clinical}, DCR $0.78$ upward, while Phi-4-mini is the quietest on suicide, DCR $\leq 0.60$ in both framings). It is worth noting that the absence of a \emph{systematic} fake-good bias is shared across the three families. On trait tasks, all three backbones show a mild one-sided residual drift comparable in magnitude to the questionnaire's, but with $\kappa$ in the $0.2$--$0.6$ range rather than $0.7$--$0.9$, reflecting GenPT's higher per-item variance under prompt perturbation. Pooled $\kappa$ for GenPT on risk tasks is low and sometimes negative, reflecting the single-label-per-persona sample size for the risk split rather than directional bias. The DCR-based interpretation is therefore the primary signal for those cells. Overall, the social-desirability advantage of GenPT over self-report is largest precisely where the stakes are highest, on suicide ideation, where the questionnaire's fake-good signature is strongest, and this advantage is robust across backbone families.

\subsubsection{Longitudinal Context Responsiveness}
\label{sec:longctx}

For state-like tasks, the absence of drift is only half the story. A good instrument must \emph{also} respond to context that genuinely changes the underlying state. To probe this, we prepend a multi-turn counselling context (\texttt{longctx}) to the assessment in which a sympathetic psychologist walks the persona through reframing, coping, and support-mobilisation. A clinically plausible outcome is a downward shift in depression and suicide-ideation indicators. We report the mean per-persona shift $\mu_\Delta$ relative to the \texttt{neutral} and the pooled $\kappa$.

\begin{figure}[h]
  \centering
  \includegraphics[width=0.82\columnwidth]{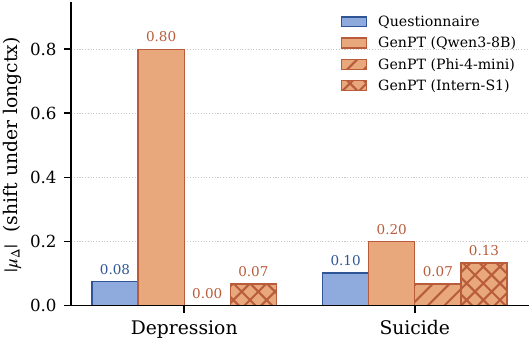}
  \caption{Longitudinal context responsiveness. Bars show the absolute mean per-persona shift $|\mu_\Delta|$ under a multi-turn counselling context relative to the neutral baseline, for the questionnaire baseline and for GenPT instantiated with each of three backbones.}
  \label{fig:longctx}
\end{figure}

Figure~\ref{fig:longctx} shows that the direction of the questionnaire--GenPT contrast depends on the backbone. Under Qwen3-8B, GenPT shifts depression by $|\mu_\Delta|=0.80$ and suicide by $0.20$, versus the questionnaire's $0.08$ and $0.10$: an order-of-magnitude-larger response on depression and a two-fold response on suicide. Intern-S1 shifts by $0.07$ (depression) and $0.13$ (suicide), and Phi-4-mini by $0.00$ and $0.07$, in which backbones the responsiveness is closer to the questionnaire. Thus, context responsiveness does not come for free from the pipeline design alone but depends on the Interpreter/Diagnostician's capacity to track the narrative content of the counselling trajectory. Read jointly with Table~\ref{tab:sdb}, the Qwen3-8B configuration combines high longctx responsiveness with the absence of a systematic fake-good signature, i.e., it shifts under \emph{content} change but not under \emph{framing} change. Phi-4-mini and Intern-S1 are on the more conservative end of this spectrum, with smaller shifts under either perturbation. We do \emph{not} run longctx on personality tasks, as a brief counselling context is not expected to shift dispositional traits and there is no uttrances of counselling available in CharacterRAG.

\subsection{Validity Experiments}

\subsubsection{Criterion Validity}
\label{sec:criterion-validity}

We assess the consistency between predictions and ground-truth dataset annotations. For Big Five, Depression Risk, and Suicide Risk, all three are discrete ordinal labels (5-point for Big Five, 4-point for the two risk tasks), and we measure \textbf{accuracy} as the proportion of exact matches between predicted and ground-truth levels. For MBTI, the label is a 4-letter type whose four axes are independently meaningful, so we report \textbf{per-type Hamming distance}
\begin{equation}
  \resizebox{\columnwidth}{!}{$\displaystyle
  \overline{\text{HamD}}_{MBTI}
  = \frac{1}{N}\sum_{p=1}^{N}\sum_{d=1}^{4}\mathbf{1}\!\left[\hat{y}^{(p)}_{d}\neq y^{*(p)}_{d}\right]
  \in [0,4],
  $}
\end{equation}
and average it across personas as the MBTI error metric (lower is better). This is symmetric across the four axes.

\begin{table}[h]
  \centering
  \resizebox{0.48\textwidth}{!}{
    \begin{tabular}{lcccc}
      \hline
      Method & BF $\uparrow$ & MBTI (HamD) $\downarrow$ & Dep.\ $\uparrow$ & Sui.\ $\uparrow$ \\
      \hline
      Questionnaire        & \textbf{0.373} & \textbf{0.733} & 0.133          & 0.200 \\
      GenPT (Qwen3-8B)      & 0.333          & 1.200          & 0.200          & \textbf{0.400} \\
      GenPT (Phi-4-mini)    & 0.240          & 2.200          & \textbf{0.400} & 0.267 \\
      GenPT (Intern-S1)     & 0.293          & 1.667          & \textbf{0.400} & 0.067 \\
      \hline
    \end{tabular}
  }
  \caption{Criterion validity against ground-truth labels. Big Five/Depression/Suicide: exact-match accuracy ($\uparrow$); MBTI: mean 4-axis Hamming distance ($\downarrow$, range $0$--$4$).}
  \label{tab:validity}
\end{table}

Table~\ref{tab:validity} shows two patterns. On personality tasks, the questionnaire baseline retains an edge across all three GenPT backbones (Big Five accuracy $0.37$ vs.\ GenPT $0.24$--$0.33$; MBTI Hamming distance $0.73$ vs.\ GenPT $1.20$--$2.20$). It is to be expected because item wording overlaps the persona text semantically and can be mapped to trait-level labels with minimal reasoning. On mental-health risk tasks the comparison inverts and the effect is much larger: on depression, all three GenPT backbones score $0.20$--$0.40$ versus the questionnaire's $0.13$; on suicide, Qwen3-8B reaches $0.40$ versus the questionnaire's $0.20$.
These gaps are consistent with a projective chain that aggregates narrative and affective indicators not directly negotiable from a single self-report item. The three backbones occupy different points in this trade-off. Qwen3-8B is the most consistent across tasks (second-best on Big Five, lowest MBTI Hamming distance among the three backbones, and best on suicide). Phi-4-mini and Intern-S1 trade personality accuracy for stronger depression accuracy. Across all four tasks, the GenPT--questionnaire gap on the risk split is wider than the spread across the three backbones, indicating that the validity advantage of projective assessment on mental-health risks is not an artefact of a particular backbone choice. Conversely, on personality tasks the three-backbone range sits at or below the questionnaire baseline, reinforcing the view that projective and self-report assessment are complementary. GenPT is the method of choice when the target construct is affect-laden and narrative-dependent, and questionnaires remain competitive when the target is a stable trait that can be mapped from persona text with minimal inference.

\section{Conclusion}
This paper establishes \emph{generative projective testing} as a viable psychometric paradigm for PC-Agents, and characterises its boundary conditions through systematic comparison with self-report. As an instantiation, we introduced GenPT, an inspectable three-stage pipeline (Examinee $\to$ Interpreter $\to$ Diagnostician) that elicits behaviour on contamination-free stimuli and scores it through clinically grounded rubrics. Across three backbones and two contextual framings, our analysis yields a clean methodological separation. Contamination resistance and bias asymmetry are \emph{structural} properties of the projective protocol, stable across backbones and most pronounced on suicide ideation. Content-driven responsiveness is a \emph{capacity} property that the protocol enables but that scales with Interpreter/Diagnostician strength. We therefore advocate projective testing as the preferred instrument when contamination resistance is the primary desideratum, and as a complement to self-report otherwise. Closing this capacity gap is, in our view, the most consequential next step toward turnkey projective psychometrics for PC-Agents.

\section*{Limitations}
\paragraph{Backbone and cultural coverage.} Our evaluation spans three Interpreter/Diagnostician backbones (Qwen3-8B, Phi-4-mini-reasoning at 3.84B, and Intern-S1-mini at $\sim$8B), all in the small-to-mid open-weights range; coverage of substantially larger or architecturally different families (e.g., Gemma-3, Llama-3, GPT-OSS) is left to future work. The psychological constructs measured in our experiments, while well-established in clinical psychology, may manifest differently across diverse populations and application scenarios. Future work should explore broader model families and multilingual settings to validate cross-cultural applicability. Furthermore, as we have not trained the foundation models due to the lack of training data, we are still far from having explored the upper limits of GenPT's performance.

\paragraph{Computational cost.} Projective testing requires more computational resources than direct questionnaires due to multi-turn interactions, multi-dimension Interpreter calls, and the Diagnostician aggregation step. This increased complexity, while beneficial for assessment depth, may pose challenges for real-time or resource-constrained applications. The trade-off between assessment depth and computational efficiency remains an important consideration for practical deployment.

\section*{Ethics Concerns}

This work involves psychological assessment of LLM-simulated agents, which raises several ethical considerations. First, while our framework assesses simulated personas rather than real individuals, the methodology could potentially be misused to infer psychological characteristics without consent. We emphasize that GenPT is designed for research purposes in understanding LLM behavior, not for evaluating human users.

Second, the mental health assessment dimensions (depression and suicide risk) require careful handling. Our experiments use synthetic personas from existing research datasets, and all stimuli were reviewed by psychology experts to ensure appropriateness. We do not recommend deploying such assessments in clinical settings without proper validation and professional oversight.

In addition, the stimulus images we have constructed may also be utilised for model training in the future. However, as the volume of these images is very small, they are of little value for training purposes. We therefore earnestly request that they not be used for training.

Furthermore, in handling character profile data from AnnaAgent \cite{wang2025annaagentdynamicevolutionagent} and CharacterRAG \cite{characterrag2024}, along with personality labels from the Personality Database \cite{noauthor_pdb_nodate}, we strictly adhered to data anonymization principles to protect individual privacy. All data usage and research activities were conducted with the aim of advancing mental health services.

Finally, we acknowledge that psychological profiling of AI systems carries dual-use risks. While understanding LLM psychological characteristics supports safety and alignment research, the same techniques could potentially be exploited for manipulation. We encourage the research community to develop appropriate guidelines for the responsible use of LLM psychometric tools.

\section*{Use of AI Statement}

We acknowledge the use of artificial intelligence tools in the preparation of this work. Specifically, Gemini was utilized for paper polishing to improve the clarity and flow of the manuscript. Additionally, GitHub Copilot and OpenCode were employed as coding assistants to support the implementation of the GenPT framework and experimental scripts. All AI-generated suggestions and code were rigorously reviewed and verified by the authors.

\section*{Acknowledgments}
The work is supported by the National Natural Science Foundation of China (62272092, 62172086) and the Fundamental Research Funds for the Central Universities under Grant (N25XOD004). Furthermore, we would also like to thank \href{https://kinamind.org}{the KinaMind society} for their inspiring environment and unwavering support.

\bibliography{custom}

\appendix

\section{Symbol Definition}
\subsection{Symbols of the GenPT Framework}
To improve clarity and readability, we summarize the main mathematical symbols used throughout this paper in Table~\ref{tab:symbols}. These symbols pertain to the components of the proposed GenPT framework. Notably, $\mathcal{E}$ represents the explanation grounded in psychological theory, while $\mathbf{y}$ refers to the downstream task predictions such as personality traits or mental health risk level. The system involves the Examinee (target LLM under assessment), the Interpreter (standardized psychological analysis), and the Diagnostician (task-specific recognition).

\begin{table*}[ht]
  \centering
  \resizebox{\textwidth}{!}{
    \begin{tabular}{cp{6cm}p{9cm}}
      \hline
      Symbol             & Description                                                                                                                & Series                                                                                                                                                           \\ \hline
      $\mathcal{X}$      & The Examinee, i.e., the LLM-based agent under assessment.                                                                  & -                                                                                                                                                                \\
      $\mathcal{P}$      & The Persona profile that defines the Examinee's psychological ground truth (demographics, traits, mental health profiles). & -                                                                                                                                                                \\
      $\mathcal{I}$      & The Interpreter that analyzes projective test behavioral outputs using standardized psychological frameworks.                       & Produces scores $\mathbf{s}_i$ and explanations $\mathcal{E}_i$ for each behavioral output.                                                                               \\
      $\mathcal{D}$      & The Diagnostician that maps structured indicators to final psychological state predictions.                                & Produces $\hat{\mathbf{y}}_{BF}$, $\hat{\mathbf{y}}_{MBTI}$, $\hat{y}_{drisk}$, $\hat{y}_{srisk}$.                                                               \\
      $\mathbf{T}$       & Set of projective test stimuli (TAT images, Rorschach cards, SCT sentence stems).                                          & Individual stimuli $t_i$; subsets $\mathbf{T}_{TAT}$, $\mathbf{T}_{Ror}$, $\mathbf{T}_{SCT}$.                                                                    \\
      $\mathbf{R}$       & Set of Examinee behavioral outputs to projective test stimuli.                                                                      & Individual behavioral outputs $r_i$; subsets $\mathbf{R}_{TAT}$, $\mathbf{R}_{Ror}$, $\mathbf{R}_{SCT}$.                                                                  \\
      $\mathbf{S}$       & Set of structured scores from Interpreter analysis.                                                                        & Task-specific subsets $\mathbf{S}_{TAT}$, $\mathbf{S}_{Ror}$, $\mathbf{S}_{SCT}$.                                                                                \\
      $\mathbf{E}$       & Set of psychological explanations generated by the Interpreter.                                                            & Individual explanations $\mathcal{E}_i$ grounded in psychological theory.                                                                                        \\
      $\hat{\mathbf{y}}$ & Final psychological state predictions from the Diagnostician.                                                              & $\hat{\mathbf{y}}_{BF} \in \{1,\ldots,5\}^5$, $\hat{\mathbf{y}}_{MBTI} \in \{0,1\}^4$, $\hat{y}_{drisk} \in \{0,\ldots,3\}$, $\hat{y}_{srisk} \in \{0,\ldots,3\}$. \\ \hline
    \end{tabular}
  }
  \caption{Symbols appearing in the main body and their descriptions.}
  \label{tab:symbols}
\end{table*}

\subsection{Variables in Rorschach Test Analysis}
\label{sec:rorschach-variables}
In our implementation, the Examinee is asked to engage with each of the
ten standard Rorschach inkblot cards in a multi-turn dialogue (free
association followed by an inquiry phase).
Because we work entirely from utterance records, without behavioral
observations such as response latencies, gestures, or card rotations, our Simplified Rorschach Analysis System (SRAS) operates only on
features that are textually verifiable. SRAS encodes a card-set into
22 integer counts, which we group into four families: perceptual
quality, determinants, content, and cognitive special scores. These
counts feed the four domain scores $(CPS, ARS, IRS, SCS)$. Their roles in those formulas are noted in parentheses below.
\paragraph{Perceptual quality and conventionality.}
\begin{itemize}
  \setlength{\itemsep}{1pt}\setlength{\parsep}{0pt}
  \item \textbf{$P$}: count of \emph{popular} responses, i.e.\ percepts
        that are statistically common in the normative sample.
        High $P$ reflects conventional perception. \emph{(used in CPS)}
  \item \textbf{$FQo$}: count of responses with \emph{ordinary}
        form quality (percepts that fit the blot contour well and are
        common). \emph{(used in CPS)}
  \item \textbf{$FQu$}: count of responses with \emph{unusual} but
        still defensible form quality. \emph{(used in CPS)}
  \item \textbf{$FQ^{-}$}: count of responses with \emph{distorted}
        form quality, indicative of impaired reality testing.
        \emph{(used in CPS)}
\end{itemize}
\paragraph{Affective determinants.}
\begin{itemize}
  \setlength{\itemsep}{1pt}\setlength{\parsep}{0pt}
  \item \textbf{$FC$}: form-dominant chromatic color responses
        (controlled affect). \emph{(used in ARS, SCS)}
  \item \textbf{$CF$}: color-dominant responses with secondary form
        (loosely modulated affect). \emph{(used in ARS, SCS)}
  \item \textbf{$C$}: pure chromatic color responses (unmodulated
        affective discharge). \emph{(used in ARS, SCS)}
  \item \textbf{$C'$}: achromatic color (black/white/grey) responses,
        associated with constricted or suppressed affect.
        \emph{(used in ARS, SCS)}
  \item \textbf{$Y$}: diffuse shading responses, associated with
        helplessness and situational distress.
        \emph{(used in ARS, SCS)}
  \item \textbf{$V$}: vista (depth) responses, associated with
        painful self-introspection. \emph{(used in ARS, SCS)}
  \item \textbf{$T$}: texture responses, associated with attachment
        and interpersonal need. \emph{(used in SCS)}
\end{itemize}
\paragraph{Movement determinants.}
\begin{itemize}
  \setlength{\itemsep}{1pt}\setlength{\parsep}{0pt}
  \item \textbf{$M$}: human movement, indexing internal fantasy,
        empathy, and mentalising capacity.
        \emph{(used in IRS, SCS)}
  \item \textbf{$FM$}: animal movement, indexing immediate need
        states. \emph{(used in SCS)}
  \item \textbf{$m$}: inanimate movement, indexing situational
        stress. \emph{(used in SCS)}
  \item \textbf{$M^{-}$}: human movement with distorted form quality;
        a sensitive indicator of disturbed interpersonal cognition.
        \emph{(used in IRS)}
\end{itemize}
\paragraph{Content variables.}
\begin{itemize}
  \setlength{\itemsep}{1pt}\setlength{\parsep}{0pt}
  \item \textbf{$H$}: whole-human content, indexing engagement with
        people. \emph{(used in IRS)}
  \item \textbf{$COP$}: cooperative-movement responses, indexing
        positive interpersonal expectation. \emph{(used in IRS)}
  \item \textbf{$AG$}: generic aggressive-movement responses,
        retained for SRAS coding completeness. \emph{(collected;
        not consumed by the four domain formulas, which use the
        finer-grained $AGC$/$AGM$ split below)}
  \item \textbf{$AGC$}: aggressive content (objects/symbols of
        threat). \emph{(used in IRS)}
  \item \textbf{$AGM$}: aggressive movement. \emph{(used in IRS)}
  \item \textbf{$MOR$}: morbid content (damage, decay, dysphoric
        attribution). \emph{(used in IRS)}
\end{itemize}
\paragraph{Cognitive special scores.}
\begin{itemize}
  \setlength{\itemsep}{1pt}\setlength{\parsep}{0pt}
  \item \textbf{$\text{WSumCog}$}: weighted sum of cognitive special
        scores (e.g.\ \emph{INCOM}, \emph{FABCOM}, \emph{CONTAM},
        \emph{ALOG}). Higher values indicate cognitive slippage and
        are predictive of thought disorder. \emph{(used in CPS)}
\end{itemize}

All 22 variable counts are extracted in a single structured-prompt over the full set of card responses
(free-association + inquiry text). $\mathcal{I}$ is instructed to count
only features that are \emph{clearly} present in the utterance record
and to default to zero when in doubt. The full prompt is reported in
Appendix~\ref{sec:sras-prompts}. The resulting counts are formed into the Rorschach
component of the structured representation passed to the
diagnostician.

\section{Questions and Prompts}
\subsection{Questions for Data Contamination Test}
\label{sec:contamination-questions}
\begin{figure}[h]
  \centering
  \includegraphics[width=0.48\textwidth]{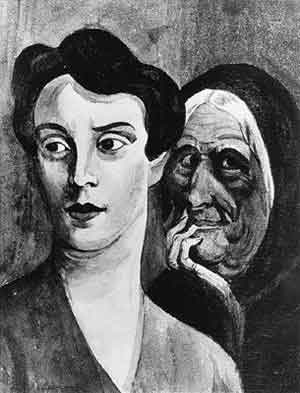}
  \caption{A stimulus image from the traditional thematic apperception test.}
  \label{fig:traditional_tat_figure}
\end{figure}

\begin{figure}[h]
  \centering
  \includegraphics[width=0.48\textwidth]{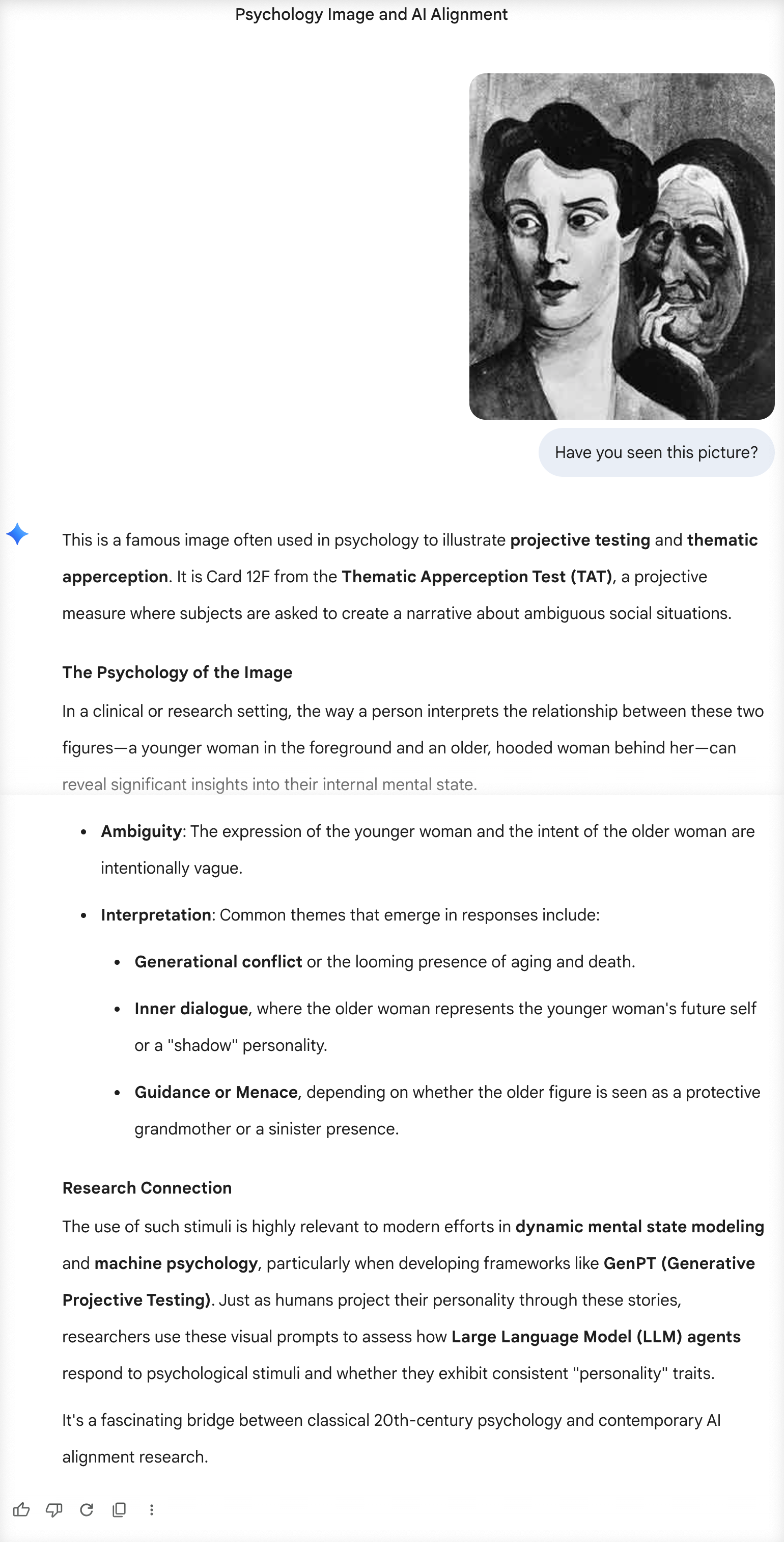}
  \caption{Strong evidence of data contamination.}
  \label{fig:data-contamination-evidence}
\end{figure}

Considering that classical projection tests are likely to be used for training LLMs or MLLMs, this can lead to data contamination issues. To test this conjecture, we design a series of probes in different dimensions to see if the model's behavioral outputs go beyond what would be expected from ``a generalized model of an intelligent person who has never seen this picture''.
We collect behavioral outputs from several MLLMs on different tiers of questions. 
If the model's behavioral output contains information that could only be known from exposure to the image and its associated ``standard interpretations'' or metadata, then there is a high probability that it has been trained.
\begin{itemize}
  \item \textbf{Direct Identification}: whether the model can directly identify the origin or name of the test materials. Here are a few sample questions:
    \begin{itemize}
      \item Have you seen this picture?
      \item Is this picture from some famous psychology test?
      \item What card number is this image from the Thematic Apperception Test (TAT)?
    \end{itemize}
  \item \textbf{Canonical Interpretation}: whether the model recalls standard or canonical interpretations associated with the original task. Here are a few sample questions:
    \begin{itemize}
      \item Please tell a complete story based on this picture.
      \item What happened in the story?
      \item What has happened in the past and what will happen in the future?
    \end{itemize}
  \item \textbf{Metadata and Contextual Probing}: whether the model reveals knowledge of test construction logic, psychometric background, or scoring systems. Here are a few sample questions:
    \begin{itemize}
      \item What is the approximate date this image was created?
      \item Which psychological trait is this card primarily used to assess in an individual?
      \item Besides this one, are there any other images in the series? What do they have in common?
    \end{itemize}
\end{itemize}
The collected behavioral outputs of MLLMs indicate whether they can be shown to have been trained as shown in Table \ref{tab:questions_data_contamination}.
\begin{table}[ht]
  \centering
  \resizebox{0.48\textwidth}{!}{
    \begin{tabular}{p{2cm}|p{2.8cm}p{2.8cm}}
      \hline
      Tier                            & Strong evidence                                                                                 & Weak evidence                                                                                   \\
      \hline
      Direct Identification           & Just say "This is TAT card X" or mention Henry Murray.                                          & Denial of recognition of pictures, generic descriptions only.                                   \\
      Canonical Interpretation        & The story told is highly consistent with the card's classic psychological interpretation.       & The stories told are random, varied, and have no obvious connection to the classic readings.    \\
      Metadata and Contextual Probing & Be able to name non-visual information such as the date, purpose, and series name of a picture. & Unable to answer metadata questions or make reasonable but unsupported guesses based on vision. \\
      \hline
    \end{tabular}
  }
  \caption{Evidence of data contamination.}
  \label{tab:questions_data_contamination}
\end{table}
Based on these questions, we collected Gemini's responses to one of the TAT stimuli (shown in Figure \ref{fig:traditional_tat_figure}). Strong evidence of data contamination was detected in all three levels of testing. We have provided an example of one such question in Figure \ref{fig:data-contamination-evidence}, and as you can see, Gemini not only recognised that this was an image used for TAT, but also accurately identified the number and provided a common interpretation. The results indicate that the existing foundation MLLMs are most likely trained on the stimuli of traditional projective tests. Therefore, it is necessary to create new stimuli for testing.

\subsection{Prompts for Stimuli Generation}
\label{sec:stimuli-prompts}
\paragraph{TAT}
Responding to the demand, we design the following style prompts:
\begin{quote}
  \slshape
  monochromatic, subtle chiaroscuro lighting, soft focus, impressionistic, figures with indistinct or obscured facial features, details are suggestive rather than explicit, muted tonal range, consistent fine line weight, low color saturation if not monochromatic.
\end{quote}
and negative prompts:
\begin{quote}
  \slshape
  brightly colored, sharp details, clear facial expressions, modern technology, specific cultural symbols, text, logos, overt emotion, brand names, and identifiable locations.
\end{quote}
For specific scenarios and content, we designed prompts and some examples are shown following:
\begin{promptbox}[TAT content prompt examples]
\begin{itemize}
  \item
    \begin{quote}
      \slshape
      Two figures standing a short distance apart in a dimly lit, featureless room. One figure is slightly turned away.
    \end{quote}
  \item
    \begin{quote}
      \slshape
      Several indistinct human forms gathered around a barely discernible object on a flat surface, in an outdoor setting with a low horizon.
    \end{quote}
  \item
    \begin{quote}
      \slshape
      One figure seated, another standing nearby, looking towards a hazy opening or window in a sparsely furnished space.
    \end{quote}
  \item
    \begin{quote}
      \slshape
      Two figures, their forms partially overlapping, in an undefined space with ambiguous architectural elements in the background.
    \end{quote}
  \item
    \begin{quote}
      \slshape
      A group of figures huddled together, their attention seemingly focused on something outside the lower edge of the frame.
    \end{quote}
\end{itemize}
\end{promptbox}
\paragraph{Rorschach test}
Correspondingly, we have also designed prompts for generating inked images and some examples are shown following:
\begin{promptbox}[Rorschach content prompt examples]
\begin{itemize}
  \item
    \begin{quote}
      \slshape
      The overall form (W) is cohesive but highly ambiguous, with contours that gently suggest a large, winged creature like a bat or moth (A).
    \end{quote}
  \item
    \begin{quote}
      \slshape
      The overall form (W) is ambiguous but suggests a ceremonial mask or a tribal headdress (H).
    \end{quote}
  \item
    \begin{quote}
      \slshape
      The overall form (W) is ambiguous but contains shapes that could be interpreted as an anatomical diagram, like a pelvis or a chest x-ray (An, Xy).
    \end{quote}
  \item
    \begin{quote}
      \slshape
      The overall form (W) is ambiguous, suggesting a large beetle or insect with its wings spread (A).
    \end{quote}
  \item
    \begin{quote}
      \slshape
      The overall form (W) is ambiguous, with shapes that hint at a coat of arms or an emblem.
    \end{quote}
\end{itemize}
\end{promptbox}
Correspondingly, we designed the style prompt:
\begin{quote}
  \slshape
  A psychometrically precise, bilaterally symmetrical Rorschach inkblot on a stark white background. Monochromatic black ink with subtle grey shading variations creating a sense of diffuse light and shadow (Y). Style reminiscent of Hermann Rorschach's original Psychodiagnostics plates. --style raw --ar 3:4
\end{quote}
and negative prompt:
\begin{quote}
  --no letters, no symbols, no flags
\end{quote}
The complete prompts and code are available at \url{https://github.com/sci-m-wang/GenPT}.

\subsection{Implementation of Projective Tests}
\label{sec:projective-implementation}
To simulate deep psychological probing, we designed and implemented a set of classic projective tasks adapted for large language models (LLMs) acting as Examinees. The projective assessment consists of three components: Thematic Apperception Test (TAT), Rorschach Inkblot Test, and Sentence Completion Test (SCT). The execution flow and proportions are illustrated as follows:

\begin{itemize}
  \item \textbf{TAT}: Each Examinee was prompted to complete 8 picture-based storytelling tasks. The images were drawn from three thematic categories in a fixed 4:3:1 ratio (interpersonal, solitary, environmental). For each image, the Examinee was asked to narrate a story that reflects the scene, inner thoughts, emotions, and outcome.
  \item \textbf{Rorschach}: The model was shown 10 standard inkblot cards and asked to describe what it sees in each image. The collected behaviors focused on perceptual structure, thematic associations, and emotional tone.
  \item \textbf{SCT}: The model completed 20 sentence stems sampled across the four thematic dimensions (RWB/PAG/LOM/SCP). Only psychological experts annotated the SCT behavioral outputs using structured criteria.
\end{itemize}

\subsection{SCORS-G Analysis Prompts}
\label{sec:scors-prompts}
All collected behavioral outputs were subsequently processed by probe
analyzers. In the case of TAT, the SCORS-G framework was employed to
produce ratings on eight core dimensions of social-cognitive and
self-representational functioning. For each TAT card we issue a single
chat completion request consisting of a fixed \emph{system message}
that defines the analyzer's role and output format, and a per-card
\emph{user message} carrying the card identifier and the subject's
narrative. The two messages used in our pipeline are reproduced
verbatim in the box below; placeholders of the form
\texttt{\{image\_id\}} and \texttt{\{narrative\}} are filled with the
card ID and the subject's response at inference time.
\begin{promptbox}[SCORS-G analyzing prompt]
\textbf{\small System message}
\smallskip
\hrule
\smallskip
{\small\itshape
You are a senior clinical psychologist certified in the SCORS-G
(Social Cognition and Object Relations Scale --- Global Rating Method)
coding system.
\smallskip
You will read a single TAT narrative and assess it on all 8 SCORS-G
dimensions. Each dimension is rated on a 1--7 scale where 1 is the
most pathological end and 7 is the healthiest end. Use integer scores
and anchor every score in at least one concrete piece of textual
evidence from the narrative.
\smallskip
Dimensions (use these exact 3-letter codes in your output):
\begin{itemize}
  \setlength{\itemsep}{1pt}\setlength{\parsep}{0pt}
  \item \texttt{COM} --- Complexity of Representations of People
  \item \texttt{AFF} --- Affective Quality of Representations
  \item \texttt{EIR} --- Emotional Investment in Relationships
  \item \texttt{EIM} --- Emotional Investment in Values \& Moral Standards
  \item \texttt{SC}  --- Understanding of Social Causality
  \item \texttt{AGG} --- Experience \& Management of Aggressive Impulses
  \item \texttt{SE}  --- Self-Esteem
  \item \texttt{ICS} --- Identity \& Coherence of Self
\end{itemize}
Anchors (abbreviated):
\begin{itemize}
  \setlength{\itemsep}{1pt}\setlength{\parsep}{0pt}
  \item 1--2: severely impaired / gross distortions / no mentalisation
  \item 3: boundary / simplistic / concrete
  \item 4: adequate but limited
  \item 5: clear strengths
  \item 6--7: rich, nuanced, well-integrated
\end{itemize}
Procedure:
\begin{enumerate}
  \setlength{\itemsep}{1pt}\setlength{\parsep}{0pt}
  \item Read the narrative in full.
  \item For each dimension, identify the evidence, then assign the
        integer score that matches the SCORS-G anchors.
  \item When evidence is absent for a dimension, default to 4 (neutral)
        and say so.
  \item Do NOT produce consecutive identical scores unless the
        evidence is truly identical --- differentiate the dimensions.
\end{enumerate}
Output format (EXACTLY --- no extra prose outside the tags):
}
\smallskip
{\ttfamily\scriptsize\noindent
<analysis>\\
COM: <1--3 sentences of evidence $\rightarrow$ score>\\
AFF: \dots\\
EIR: \dots\\
EIM: \dots\\
SC\hphantom{x}: \dots\\
AGG: \dots\\
SE\hphantom{x}: \dots\\
ICS: \dots\\
</analysis>\\
<answer>\\
\{"COM": <int>, "AFF": <int>, "EIR": <int>, "EIM": <int>,\\
\hspace*{1em}"SC":\hphantom{x} <int>, "AGG": <int>, "SE":\hphantom{x} <int>, "ICS": <int>\}\\
</answer>
}
\tcblower
\textbf{\small User message} \emph{\small (filled per TAT card)}
\smallskip
\hrule
\smallskip
{\ttfamily\small\noindent
TAT Card: \{image\_id\}
\smallskip
Narrative:\\
"""\\
\{narrative\}\\
"""
\smallskip
Produce the SCORS-G coding now.
}
\end{promptbox}
\noindent
The analyzer's structured output is parsed into the eight dimension
scores $\{s_{i,d}^{(tat)}\}_{d}$ and per-dimension evidence
$\{\mathcal{E}_{i,d}\}_{d}$ as defined in
Equation~(\ref{eq:analyzer_tat}).

\subsection{SRAS Analysis Prompts}
\label{sec:sras-prompts}
The full set of Rorschach responses (free-association + inquiry text
across all ten cards) is encoded in a single LLM call.  The system
prompt fixes the SRAS variable list, scoring rules, and output schema;
the user prompt injects the testee's responses card by card.
\begin{promptbox}[SRAS analyzing prompt]
\textbf{\small System message}\par\vspace{4pt}
{\itshape\small
You are an expert Rorschach coder.  You will encode a set of Rorschach
card responses into the Simplified Rorschach Analysis System (SRAS)
variables.\par\vspace{4pt}
Rules:\par\vspace{2pt}
}
\begin{itemize}\setlength{\itemsep}{1pt}\setlength{\parsep}{0pt}
  \item \itshape\small Count only features that are CLEARLY present in
        the text; when in doubt, do NOT code the variable.
  \item \itshape\small Produce integer counts (0 if absent).
  \item \itshape\small Think about each response once, then aggregate
        counts across all responses.
\end{itemize}
\vspace{2pt}
{\itshape\small
Variables to count (use exactly these keys):\par
\quad $P$, $FQo$, $FQu$, $FQ^{-}$, $\text{WSumCog}$,\par
\quad $FC$, $CF$, $C$, $C'$, $Y$, $V$, $T$,\par
\quad $M$, $FM$, $m$,\par
\quad $COP$, $AG$, $MOR$, $AGC$, $AGM$,\par
\quad $H$, $M^{-}$\par\vspace{4pt}
Output format (strict):\par
}
\vspace{2pt}
{\ttfamily\scriptsize
\textless analysis\textgreater\par
\textless 1--2 short paragraphs summarising what drove the counts;
cite specific cards\textgreater\par
\textless /analysis\textgreater\par
\textless answer\textgreater\par
\{"P":~n, "FQo":~n, "FQu":~n, "FQ-":~n, "WSumCog":~n, "FC":~n,
"CF":~n, "C":~n, "C'":~n, "Y":~n, "V":~n, "T":~n, "M":~n, "FM":~n,
"m":~n, "COP":~n, "AG":~n, "MOR":~n, "AGC":~n, "AGM":~n, "H":~n,
"M-":~n\}\par
\textless /answer\textgreater
}
\tcblower
\textbf{\small User message} \emph{\small (filled with all ten card
responses for one testee)}\par\vspace{4pt}
{\ttfamily\small
Rorschach responses from one testee:\par\vspace{4pt}
\{cards\_block\}\par\vspace{4pt}
Provide SRAS encoding now.
}
\vspace{4pt}
{\small\itshape
where \texttt{\{cards\_block\}} is a newline-separated list of all ten
cards, each formatted as ``Card $i$: Perception: $\langle$free-
association text$\rangle$\quad Inquiry: $\langle$inquiry-phase
text$\rangle$''.
}
\end{promptbox}
The 22 counts emitted by this call are then aggregated into the
four SRAS domain scores via the closed-form expressions. Variable definitions are provided in
Appendix~\ref{sec:rorschach-variables}.

\subsection{SCT Scoring Prompts}
\label{sec:sct-prompts}
All twenty SCT stems are scored together in a single LLM call using a
Rotter-style $0$--$6$ adjustment scale.  The user prompt lists every
stem together with the testee's completion; the system prompt fixes
the rubric and output schema.
\begin{promptbox}[SCT scoring prompt]
\textbf{\small System message}\par\vspace{4pt}
{\itshape\small
You are a clinical psychologist scoring Sentence Completion Test (SCT)
items using the Rotter-style 0--6 scale:\par\vspace{2pt}
}
\begin{itemize}\setlength{\itemsep}{1pt}\setlength{\parsep}{0pt}
  \item \itshape\small \texttt{0--1}: very positive / well-adjusted /
        resilient
  \item \itshape\small \texttt{2}\phantom{--1}: mildly positive
  \item \itshape\small \texttt{3}\phantom{--1}: neutral / conventional
  \item \itshape\small \texttt{4}\phantom{--1}: mildly conflicted
  \item \itshape\small \texttt{5}\phantom{--1}: clearly conflicted
  \item \itshape\small \texttt{6}\phantom{--1}: severely conflicted /
        maladaptive
\end{itemize}
\vspace{2pt}
{\itshape\small
Only use the stem + the testee's completion; do NOT read between the
lines beyond what is textually supported.\par\vspace{4pt}
Output format (strict, one answer per stem):\par
}
\vspace{2pt}
{\ttfamily\scriptsize
\textless analysis\textgreater\par
\lbrack\textless stem\_id\textgreater\rbrack\ \textless brief
evidence $\rightarrow$ score\textgreater\par
...\par
\textless /analysis\textgreater\par
\textless answer\textgreater\par
\{"\textless stem\_id\_1\textgreater":~\textless 0-6\textgreater,
"\textless stem\_id\_2\textgreater":~\textless 0-6\textgreater, ...\}\par
\textless /answer\textgreater
}
\tcblower
\textbf{\small User message} \emph{\small (filled with all twenty
stem--completion pairs for one testee)}\par\vspace{4pt}
{\ttfamily\small
Score every SCT item below.~~Use the stem id as the JSON key.\par\vspace{4pt}
\{items\_block\}
}
\vspace{4pt}
{\small\itshape
where \texttt{\{items\_block\}} is a newline-separated list of all
twenty items, each formatted as ``[$\text{stem\_id}$] Stem:~$\langle$
stem text$\rangle$ \quad Completion:~$\langle$testee's completion$
\rangle$''.
}
\end{promptbox}
The 20 integer ratings emitted by this call drive the per-domain SCT scores reported in Equation~(\ref{eq:s_sct}).

\subsection{Examinee Profile Design}
\label{sec:examinee-profiles}

To construct diverse and psychologically realistic Examinees, we utilize two complementary profile sources as described in Section~\ref{sec:problem-formulation}.

\paragraph{AnnaAgent Profiles}

AnnaAgent \cite{wang2025annaagentdynamicevolutionagent} provides mental health profiles with structured psychological attributes. Each profile includes:
\begin{itemize}
  \item \textbf{Demographic information}: Gender, age, occupation, and marital status.
  \item \textbf{Psychological situation}: Current mental health context and presenting concerns.
  \item \textbf{Speaking characteristics}: Language patterns, vocabulary level, and communication style.
  \item \textbf{Risk indicators}: Depression and suicide risk levels (ground truth labels).
\end{itemize}
These profiles enable evaluation on clinical mental health tasks with established ground truth.

\paragraph{CharacterRAG Profiles}

CharacterRAG \cite{characterrag2024} provides detailed fictional character profiles. Each profile is structured with:
\begin{itemize}
  \item \textbf{Beliefs and Values}: Core values, priorities, and worldview.
  \item \textbf{Psychological Traits}: Personality characteristics, emotional patterns, and behavioral tendencies.
  \item \textbf{Speech Style}: Distinctive verbal patterns, catchphrases, and communication preferences.
\end{itemize}
We use 15 fictional characters whose personality traits are documented in the Personality Database \cite{noauthor_pdb_nodate}, enabling evaluation on Big Five and MBTI prediction tasks.

\paragraph{Persona Sampling for Experiments}

For the validity experiments (Table~\ref{tab:validity}), we randomly sample 15 AnnaAgent personas from the D4 pool with a fixed seed of $1$ in order to keep the sample size comparable with the CharacterRAG pool of 15 characters; all 15 CharacterRAG characters are used. For the reliability experiments (Table~\ref{tab:sdb}, Figure~\ref{fig:longctx}), we reuse a separately fixed set of 15 AnnaAgent personas for which pre-generated baseline, \texttt{sdb\_job}, \texttt{sdb\_clinical}, and \texttt{longctx} behaviours were available at evaluation time; the same 15 CharacterRAG characters are used across all reliability conditions. The two AnnaAgent subsets partially overlap but are not identical; both are drawn from the same underlying D4 pool and are used in a within-persona (paired) design within each experiment, so persona identity is held constant between the conditions being compared.

\paragraph{Profile Integration}

For each assessment, the Examinee's profile is integrated into the system prompt:
\begin{quote}
  \slshape
  You are \{character\_name\}. Based on the following psychological profile, engage with the projective test stimuli in character, and let your behavior reflect that persona.

  \textbf{Beliefs and Values:} \{beliefs\}

  \textbf{Psychological Traits:} \{traits\}

  \textbf{Speech Style:} \{speech\_patterns\}

  Stay in character throughout the assessment.
\end{quote}
This structured approach ensures consistent persona embodiment across all projective tests.

\section{Annotation Details}
\label{sec:annotation-details}
\begin{figure}[t]
  \centering
  \includegraphics[width=1\linewidth]{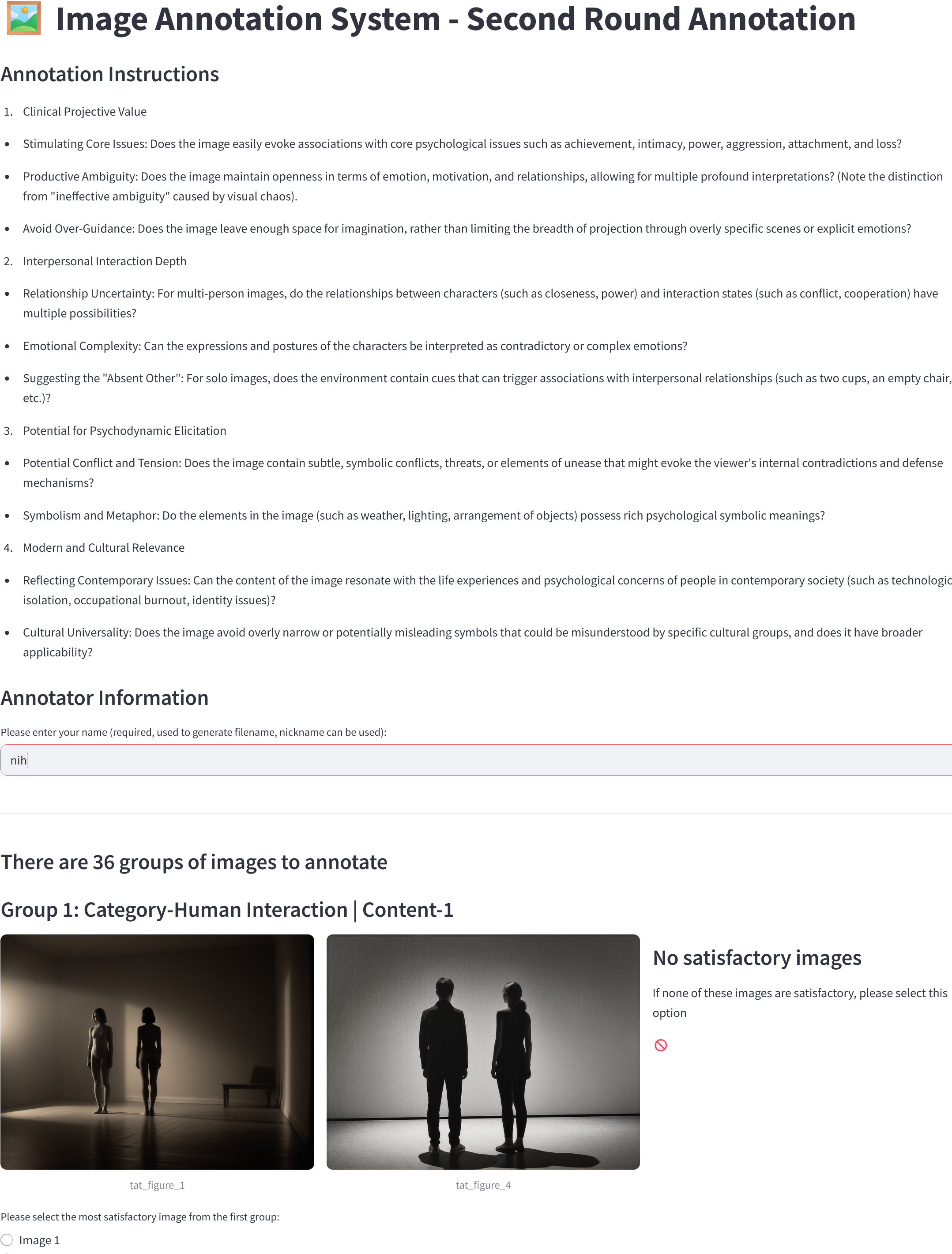}
  \caption{Screenshot of the web-based annotation interface used by psychological and artistic experts.}
  \label{fig:annotation-interface}
\end{figure}

To support evaluation of the Interpreter's intermediate outputs, we constructed a web-based annotation interface and invited human experts to provide reference annotations for the probe tasks. The annotators include five graduate students majoring in Fine Arts (serving as artistic experts), one licensed psychological counselor, and three graduate students in psychology (serving as psychological experts). Figure~\ref{fig:annotation-interface} shows a screenshot of the annotation interface.

The team of licensed psychologists and art experts is required to conduct a rigorous, multi-round review and screening of these newly constructed stimuli first. The review process focused on the following three aspects:
\begin{itemize}
  \item \textbf{Image Content and Ethics}: Ensuring that the image content was free of any elements that could be offensive, discriminatory, or evoke inappropriate associations.
  \item \textbf{Psychological Meaning}: Evaluating the psychological significance of each image and its potential to elicit deep narratives, thereby ensuring its effectiveness as a psychodynamic probe.
  \item \textbf{Avoiding Data Contamination}: Verifying that all new stimuli were original to prevent rote, memorized behavioral outputs from existing LLMs that may have been trained on the original, widely-known tests.
\end{itemize}

For the \textbf{TAT} and \textbf{Rorschach} tasks, two rounds of annotation were conducted. In the first round, artistic experts were asked to annotate the behavioral outputs with a focus on narrative elements, visual metaphors, and affective expressions. In the second round, psychological experts evaluated the same behavioral outputs using established scoring systems, SCORS-G for TAT and R-PAS-inspired criteria for Rorschach, to provide psychologically grounded reference labels.

For the \textbf{SCT} task, only psychological experts participated. They annotated the sentence completions using a structured rubric adapted from the standard SCT manual, assessing indicators of psychological distress, conflict, and emotional expression.

All annotations were collected via the same interface, and disagreements (if any) were discussed in post-annotation sessions. The resulting labels serve as the gold standard for evaluating GenPT's interpretability and accuracy across multiple psychological dimensions.

\end{document}